\documentclass{aastex62}

\newcommand\altsched{\textsc{ALTSched}\xspace}
\newcommand\minion{\textsc{minion\_1016}\xspace}
\newcommand\opsim{\textsc{OpSim}\xspace}

\graphicspath{{./}{figures/}}

\usepackage{natbib}
\usepackage{verbatim}
\usepackage{xspace}
\usepackage{mathrsfs}
\usepackage{amsmath}
\usepackage[page]{appendix}
\usepackage{float}
\usepackage{tabularx}
\usepackage{textcomp}
\newcolumntype{K}{>{\raggedright\arraybackslash}X}
\newcolumntype{E}{>{\raggedleft\arraybackslash}X}

\newcommand{\tildemid}{\raisebox{0.5ex}{\texttildelow}}

\begin{document}

\title{ALTSched: Improved Scheduling for Time-Domain Science with LSST}

\correspondingauthor{Daniel Rothchild}
\email{drothchild@berkeley.edu}

\author[0000-0002-4605-0949]{Daniel Rothchild}
\affil{Department of Electrical Engineering and Computer Science, University of California, Berkeley}

\author[0000-0003-0347-1724]{Christopher Stubbs}
\affiliation{Department of Physics, Center for Astrophysics, Harvard University}

\author[0000-0003-2874-6464]{Peter Yoachim}
\affiliation{Department of Astronomy, University of Washington}

\begin{abstract}
Telescope scheduling is the task of determining the best sequence of observations (pointings and filter choices) for a survey system. Because it is computationally intractable to optimize over all possible multi-year sequences of observations, schedulers use heuristics to pick the best observation at a given time. A \textit{greedy} scheduler selects the next observation by choosing whichever one maximizes a scalar merit function, which serves as a proxy for the scientific goals of the telescope. This sort of bottom-up approach for scheduling is not guaranteed to produce a schedule for which the sum of merit over all observations is maximized. As an alternative to greedy schedulers, we introduce \altsched, which takes a top-down approach to scheduling. Instead of considering only the next observation, \altsched makes global decisions about which area of sky and which filter to observe in, and then refines these decisions into a sequence of observations taken along the meridian to maximize SNR. We implement \altsched for the Large Synoptic Survey Telescope (LSST), and show that it equals or outperforms the baseline greedy scheduler in essentially all quantitative performance metrics. Due to its simplicity, our implementation is considerably faster than \opsim, the simulated greedy scheduler currently used by the LSST Project: a full ten year survey can be simulated in 4 minutes, as opposed to tens of hours for \opsim. LSST's hardware is fixed, so improving the scheduling algorithm is one of the only remaining ways to optimize LSST's performance. We see \altsched as a prototype scheduler that gives a lower bound on the performance achievable by LSST.

\end{abstract}


\section{Introduction} \label{sec:intro}
LSST is a large ground-based survey system scheduled to begin operations in 2022.
With the capability to survey the entire southern sky in 6 bands about once a week,
LSST will enable
time-domain astronomy at an unprecedented scale. The telescope was designed
with four primary science goals in mind: to understand dark matter and dark
energy; to catalog the solar system; to study the structure and formation of
the Milky Way; and, perhaps most uniquely suited to LSST's particular
characteristics, to explore the large frontier of time-domain astronomy.
These science goals motivate a range of competing technical metrics
that LSST should optimize. Presenting a full list of science cases for LSST and accompanying
technical metrics is beyond the scope of this paper; we refer readers to the LSST observing
strategy whitepaper \citep{surveystrat}. Here we present only a few representative
examples (with science cases in parentheses), which we believe are especially
sensitive to scheduling decisions:

\begin{tabularx}{\textwidth}{EcK}
\vspace{10mm} \\
\textbf{long observing seasons} (parallax measurements, AGN variability, few-months-long transients)
& $\iff$ & 
\textbf{high cadence} (faster transients, better characterization of light curves) \\
\textbf{large survey area} (large-scale structure, weak lensing, galaxy/star surveys)
& $\iff$ & 
\textbf{higher co-added depth} (studying fainter objects) \\
\textbf{few filter changes} (maximize survey efficiency)
& $\iff$ & 
\textbf{many filter changes} (obtain nightly colors for variables \& transients) \\
\textbf{large overlap between neighboring observations} (rapid revisits for very fast transients)
& $\iff$ & 
\textbf{small overlap between neighboring observations} (cover a larger area per unit time) \\
\textbf{multiple observations per field per night} (link asteroids, probe \tildemid hour-scale variability)
& $\iff$ &
\textbf{one observation per field per night} (transients that don't change in \tildemid 1 hour) \\
\vspace{10mm}
\end{tabularx}

With the LSST Project well into the construction phase, the system's hardware
characteristics (aperture, field size, slew rate, field of view, sensor
quantum efficiency...) are fixed.  Apart from actively engineering the weather
in Chile, the only remaining opportunities to extract additional performance from
the LSST system are in the scheduler and in the data reduction pipeline.
LSST's existing scheduler, called \opsim,
is a greedy algorithm: it chooses the next field to observe by maximizing a scalar merit
function intended as a proxy for the scientific merit of observing the field. This approach
is appealing in its apparent simplicity, but as we show below, even after over a decade of
development, \opsim under-performs on many science metrics, particularly in the time-domain
science so critical to LSST's mission. We argue here that the fault lies not in any
implementation detail, but rather in the fundamental nature of greedy algorithms:
because they make scheduling decisions only one or a few observations in advance,
the parameters of the algorithm give scientists little or no control over many 10-year
global properties of the schedule.

To remedy this, we introduce and analyze an alternative scheduling algorithm, \altsched,
which makes scheduling decisions with a top-down hierarchical approach, instead of
using a purely bottom-up local merit function. We defer a full description of
\altsched to Section \ref{sec:altsched}, but simply put, our algorithm first decides
which region of sky to observe (nightly), then which particular area and filter to use
(hourly), and only then which individual pointings to observe (minute-scale).
In contrast, \opsim chooses every next observation ab initio, with no planning horizon
larger than 30 seconds and no global strategy except what is encoded indirectly through
the merit function.

One major advantage of \altsched is that the parameters
of the algorithm directly control final properties of the schedule. For example,
\altsched observes either the Northern or Southern sky each night. Because LSST
can observe about half of the sky each night, we can obtain a universal 2-day
revisit cadence simply by assigning even-numbered nights to the Northern sky and
odd-numbered nights to the Southern sky. Different cadences (e.g. with some 1- and 3-day
revisits) are equally simple with a different pattern of North and South. Importantly,
there is no known way to similarly modify this distribution of revisit times in \opsim,
except by adjusting the weights and penalties of the merit function
specifically to mimic the exact behavior of \altsched.

As of the writing of this paper, the LSST Project is planning to run a suite
of \opsim simulations in order to choose a final observing strategy for LSST.
One point of this paper is to argue that optimizing over the parameters of \opsim
in this way will not yield an optimal survey strategy. Taking the example above
one step further before giving a fuller analysis in \ref{sec:results},
the gaps between observations to a particular sky pixel resemble an exponential
distribution for every \opsim run we have analyzed, over a wide range of \opsim's parameters.
An exponential distribution of revisit times is consistent with the
timing of observations following a random process. In \altsched, the shape of this
distribution can be easily controlled. Instead of running more and more \opsim
simulations, we therefore advocate exploring how \opsim can be made to reproduce
the survey strategy of \altsched, and how \altsched's cadence can be further improved.

In the sections that follow, we first summarize the LSST baseline scheduler and
the tools used to assess the scientific performance of various alternatives.
We then describe the implementation of our alternative scheduler, \altsched,
and make quantitative performance comparisons to the LSST baseline.  We close
with a discussion and some suggestions for future work. 

\section{\opsim}
To provide context for \altsched, the algorithm we propose below, we first
introduce LSST's current baseline scheduling algorithm.\footnote{The LSST
scheduler is undergoing continual development, and the version of \opsim described
and compared with in this article has been superseded, but it was the baseline
plan at the time this work was completed.}

LSST's current scheduler is part of a package called the Operations
Simulator (\opsim) \citep{delgado2016lsst}. \opsim uses a greedy
algorithm to choose fields based on a proposal system, where
abstractions of different scientific proposals give a score for
each candidate field, and the scheduler chooses the field with
the highest combined score, or merit. \opsim considers a number of criteria
in its merit function, including time since last observation, co-added depth
achieved thus far, airmass and sky brightness at the proposed pointing, slew time
to reach the pointing, etc. For a full description of the complex merit
function used for \minion, we refer readers
to \citet[\S 5]{delgado2014lsst}. Once an observation is chosen and executed,
the merit scores are recalculated for every candidate field, and the process repeats.
The parameters of \opsim are the weights and penalties associated with the various
inputs to the merit function. Although easily interpretable in the context of local decision
making, these parameters have no simple connection to many of the scheduling decisions
we actually want to make: the average and distribution of season lengths and of revisit
times; the colors (or lack thereof) obtained during a single night; the uniformity
of light curve sampling over time; etc.
In contrast, as we show below, \altsched's parameters directly control these
properties.

Included with \opsim is a module that simulates the system hardware, weather,
seeing conditions, and downtime. For fairness of comparison, \altsched uses
the exact same calculations, though we re-implement the simulator to improve
computational efficiency. In particular, we obey all physical constraints on the
telescope, including making no observations below LSST's minimum elevation or in
the zenith avoidance area, accounting for readout time, shutter time, slew speed,
settle time, optics correction time, and filter change time, only using 5 of
LSST's 6 band-pass filters per night, etc.
Taking slew time as an example: both \opsim and \altsched use the same
slew calculation, described in \citet[\S 6]{delgado2014lsst}. In short, the
slew time is calculated assuming uniform accelerated motion of the telescope
mount and dome, plus some settle time.

To compare simulated surveys, the LSST Project has developed a useful suite of
analysis tools, the Metrics Analysis Framework (MAF), that can
evaluate candidate schedules by computing various performance
metrics of interest \citep{jones2014lsst}.
Examples include the number of type Ia supernova light
curves that meet certain criteria, the distribution of co-added depths over
the observed region, the anticipated uncertainty in parallax and proper motion
measurements. This framework is critical for the fair evaluation of simulated
surveys.

A number of \opsim runs, or simulated surveys, have been released by the Project.
At the time this work was carried out, the baseline simulated survey was
\minion, and we compare our results to those of \minion. Later versions of
\opsim have improved on a number of metrics, but all runs we have analyzed since
\minion suffer from poor time-domain performance. For example, none
of the cadences released with the recent whitepaper call recover even half as many
well-sampled SNIa without eliminating visit pairs or changing the exposure time
(where ``well-sampled'' is as defined below). And to our knowledge, no \opsim or
feature-scheduler run at all has exceeded \altsched in this metric.
Results for \minion are presented in more depth in Section \ref{sec:results}.

Several alternatives to \opsim have also been proposed. \cite{naghib2016} frame
the scheduling problem as a Markovian decision process. \cite{ridgway2015units}
proposes scheduling LSST by dividing observing into blocks that are observed
(and then reobserved) in an optimal manner. Ridgway's proposal is similar in
spirit to the algorithm developed and described in this paper.

\section{\altsched}
\label{sec:altsched}
In this section, we introduce \altsched, an alternative scheduling algorithm
for LSST. In the next section, we introduce a number of metrics, and 
demonstrate quantitatively that \altsched equals or outperforms
the existing \opsim baseline \minion on these metrics.

\subsection{\altsched Algorithm}
\altsched is a deterministic scheduling algorithm: in a given night, it does not adapt
to current weather or seeing conditions. Although in principle, adjusting a schedule
to take prevailing weather conditions into account should only improve a scheduler, we
justify our decision to use a non-adaptive algorithm in three ways: 1) maintaining a
consistent cadence -- i.e. avoiding long observation gaps -- is difficult to do if the
scheduler aggressively avoids regions with poorer observing conditions, since by
random chance, some regions will go unobserved for a long time; 2) the best seeing tends
to occur on the meridian (at minimum airmass), which is where \altsched observes anyway;
3) one strength  of \altsched is that it achieves such high performance even
without dynamically reacting to observing conditions. We only expect our
performance to improve with the judicious addition of poor-weather avoidance.

The algorithm itself is remarkably simple. Each night, it chooses whether to
observe the Northern or Southern sky based on which region has received fewer
visits so far, and then during the night it scans North and South of the
meridian, taking 30-second exposures and then slewing by approximately one
field width. LSST is on an alt/az mount, and so has a zenith-avoidance area.
To observe the region of sky that passes over zenith, we therefore
periodically scan to the East and West over that region. Each N/S scan is
repeated twice in order to obtain two observations per night, separated by
roughly 30-60 minutes. Before repeating a scan, we change filters so that
every field is visited in two bands per night. We refer readers to
the video in the supplementary material; the scanning strategy is much more
easily explained visually than in text.

The exact pointings used are drawn from a fixed tiling that is randomly
rotated by much more than a field size each night. If a pointing is not used
in a given night, it is saved for use in a later night, so that gaps do not
persist. The fixed tiling is chosen as a solution to the Thomson
charge-distribution problem for $N=3500$ \citep{thomson1904}. 

\subsection{Moon Avoidance}
\begin{figure}
    \centering
    \includegraphics[width=300px]{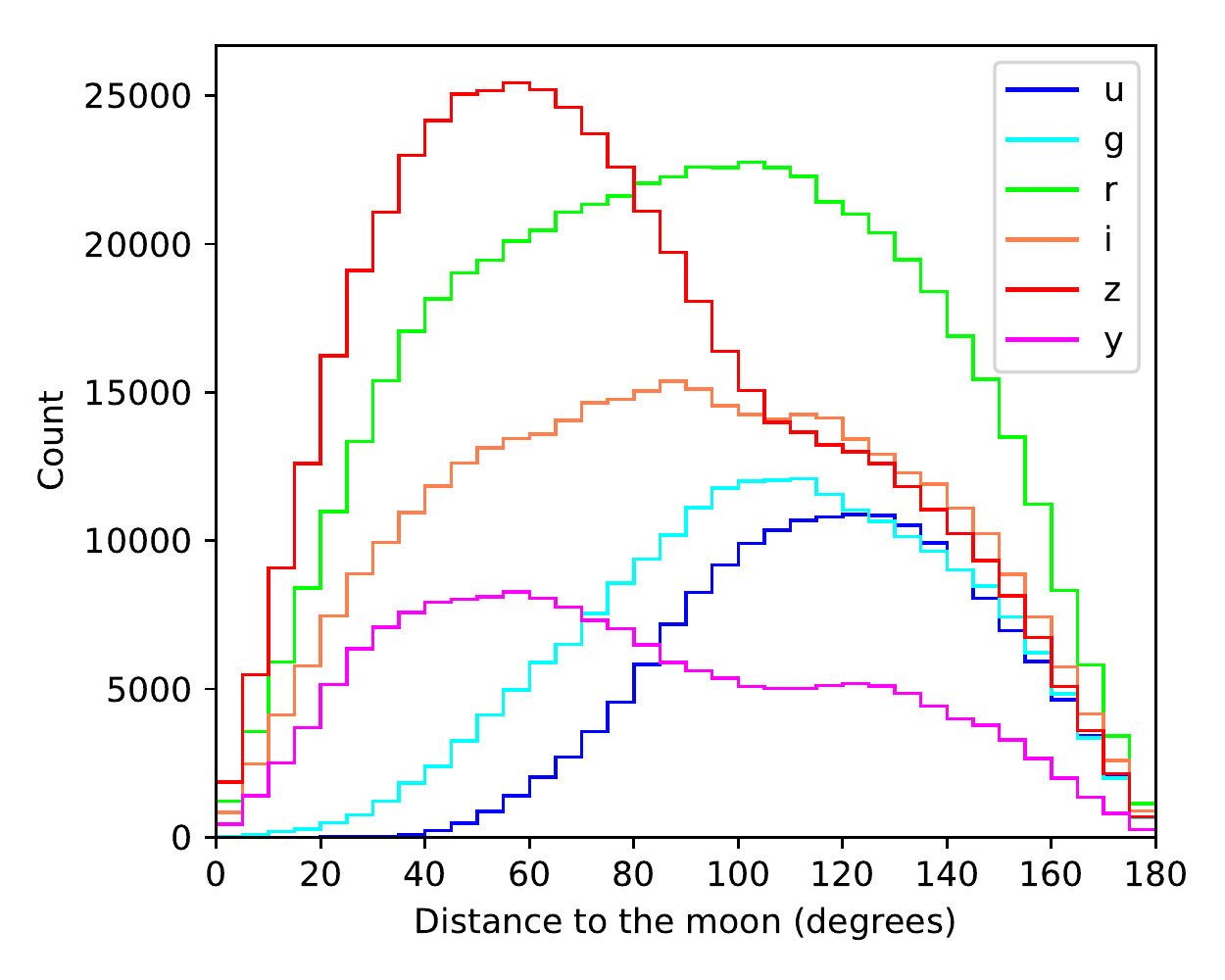}
    \caption{Histogram of angular distance between \altsched's observations and the moon.
             Only a small fraction of observations fall closer than $30^\circ$ to the moon,
             and of those, a number occur when the moon is only slightly illuminated.}
    \label{fig:moon}
\end{figure}

One deficiency in \altsched is that it contains no provision for avoiding the moon.
\minion uses a conservative avoidance radius of $30^\circ$. This corresponds to 7\%
of the celestial sphere, so about 7\% of observations should fall within
this avoidance zone. However, $30^\circ$ is likely too conservative 
in times near a new moon and in the redder filters. We therefore expect only a few
percent of observations to be problematically close to the moon.
For observations that fall too close to the moon, the simulated 
sky brightness is very high, leading to a low limiting depth for those observations.
So any metric depending on depth should only improve by adding a moon-avoidance
module to \altsched. Figure \ref{fig:moon} shows histograms of the angular distance
between \altsched's observations and the moon. Note that \altsched avoids observing
in the u and g bands at all when the moon is up and bright, so there are few or
no observations in u or g closer than $30^\circ$ to the moon.

\subsection{Filter Allocation}
Most simulated surveys of LSST assume that the telescope will visit each field twice
per night, with a separation of $\sim 30-60$ minutes, in order to link asteroid
observations. To improve the cadence in each band, \altsched usually carries out these
two visits in different filters. To accomplish this with a minimal number of filter
changes, we divide observations into blocks that take $\sim 30-60$ minutes to observe,
and we visit each block twice back-to-back, changing filters before revisiting a block
(but not between the revisit of a block and the first visit to the subsequent block).
The video in the supplementary materials demonstrates this filter allocation strategy.
On average, we execute $11.4$ filter changes per night, compared to \minion's $4.3$.
This adds $1.3$ seconds to our average slew time over \minion, and allows us to
double cadence in each band.

Our filter allocation strategy is designed such that, in theory, visits to every
sky pixel will cycle through the six filters in order. Assigning numbers 1 through
6 to LSST's 6 bandpass filters ($ugrizy$), and using arithmetic modulo 6,
we start a night in some filter $i$, and during the night, before every revisit block,
we switch from filter $k$ to filter $k+1$. The next night starts in filter
$i+1$ and the process repeats, ensuring that visits to each pixel cycle through
all six filters. In the ideal case, every sky pixel gets a visit in every band
once every 6 nights (2 bands every 2 nights, since we alternate observations between
the northern and southern sky). 

However, many factors cause us to deviate from this ideal strategy. First, we don't
want a uniform distribution of total number of visits over filters, so we replace some
observations in, say, the $y$ band, with more in $r$ instead. Second, we observe $y$
and $z$ preferentially in twilight, since these bands are less sensitive to high
sky brightness. Third, we avoid using $u$ or $g$ when the moon is up and bright.
And lastly, the filter changer can only fit 5 of LSST's 6 filters, so only 5 filters
can be used in a single night. In each of these cases, if a filter is ``not allowed''
during a time when the cyclic allocation strategy would have scheduled it, we simply use
some other filter instead.
Besides these intentional deviations, weather and downtime also cause us to deviate
from this ideal strategy in ways that are less predictable.

Overall, \altsched's filter allocation strategy ensures that the per-band cadence
is much more regular than in \minion (as shown below), but there is likely much room
for improvement by designing a strategy that more intelligently takes the four
deviations from ideal described above into account.

\section{Metrics \& Results}
\label{sec:results}
As laid out in Section \ref{sec:intro}, the science drivers of LSST motivate a
wide range of technical metrics, many of which are in tension with each other.
Broadly speaking, LSST's science goals fall into two categories: those that
depend mainly on the final co-added images (static science), and those that
depend mainly on the temporal distribution of visits throughout the survey
(time-domain science). 
A full list of results on a variety of metrics is available at
\url{http://altsched.rothchild.me:8080}. Here we highlight a small subset of these
metrics that are particularly sensitive to scheduling decisions, especially those
where \altsched and \minion achieve different results.

\subsection{Static Science}
A number of science drivers (e.g. galaxy/star surveys, large-scale structure measurements)
use co-added images, and are largely insensitive to the exact distribution of visits
over time. A wide range of static science is enabled by achieving a higher final
co-added depth in each band. We therefore use co-added depth as a chief metric
for static science. Single-visit depths and full 10-year co-added depths are shown
for both schedulers in Table \ref{table:fiducial_depths}.
Since different survey strategies may allocate visits across
filters differently, we summarize the total co-added performance across bands with the
\textit{effective survey time} metric $T_{eff}$. Given design limiting depths $M_f$ for a
30-second exposure in filter $f\in \mathscr{F}$, the effective survey time of a series
of exposures  $\mathscr{E}={(m_i, f_i)}$ which achieved a $5\sigma$ limiting depth
$m_i$ in filter $f_i$ is given by
\[T_{eff} = \sum_{(m_i, f_i)\in \mathscr{E}} 30\text{ sec.}\times 10^{0.8(m_i-M_{f_i})}.\]
The $5\sigma$ limiting depth of an exposure is computed taking into consideration
the sky brightness, seeing, and airmass, as described in equation 6 of \citet{lsst}.
Design depths $M_f$ for LSST are shown in Table \ref{table:fiducial_depths}.
These design depths assume an airmass of $1$, $r$-band seeing of $0.7$ arcsec (FWHM), and
$r$-band sky brightness of 21 mag/arcsec$^2$. In practice, observations are taken under
worse conditions, so the total $T_{eff}$ is surprisingly low for a 10-year survey.

\begin{table}
    \centering
    \begin{tabular}{lcccccc}
        & u & g & r & i & z & y \\
        \hline
        Design (Ideal) Single-Visit Depths & 23.9 & 25.0 & 24.7 & 24.0 & 23.3 & 22.1 \\
        Median Single-Visit Depth (\minion) & 23.09 & 24.51 & 24.05 & 23.45 & 22.71 & 21.78 \\
        Median Single-Visit Depth (\altsched)& 23.21& 24.50 & 23.93 & 23.49 & 22.90 & 21.91 \\
        Median Co-added Depth (\minion) &   25.48 & 27.02 & 27.03 & 26.46 & 25.65 & 24.73 \\
        Median Co-added Depth (\altsched) & 25.61 & 27.03 & 27.04 & 26.35 & 25.93 & 24.31
    \end{tabular}
    \caption{The first row shows the design specification for single-visit $5\sigma$
             limiting depths for LSST's six broad-band filters under ideal observing 
             conditions (in  magnitudes). These depths assume airmass of $1$,
             $r$-band seeing of $0.7$, arcsec (FWHM), and $r$-band sky brightness of
             21 mag/arcsec$^2$. Sunsequent rows show the actual median single-visit
             depths achieved by \minion and \altsched, and the median (over sky pixels)
             10-year co-added depths.}
    \label{table:fiducial_depths}
\end{table}

Two related metrics are the \textit{average slew time} (including filter changes), and the
\textit{open-shutter fraction}, defined as:
\[OSF = \frac{T_{exp}}{T_{exp} + T_{slew} + T_{shutter} + T_{readout}}\]
where $T_{exp}$ is the exposure time, and $T_{readout}$ consists of any intermediate
readout time between back-to-back ``snaps'' during the same visit (the readout after
the last snap is included in the slew time). Both \minion and \altsched divide
30-second visits into 2 15-second snaps for cosmic-ray rejection.

\begin{figure}
    \centerline{
        \includegraphics[width=220px]{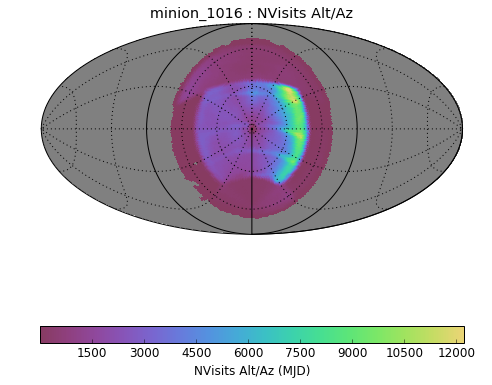}%
        \includegraphics[width=220px]{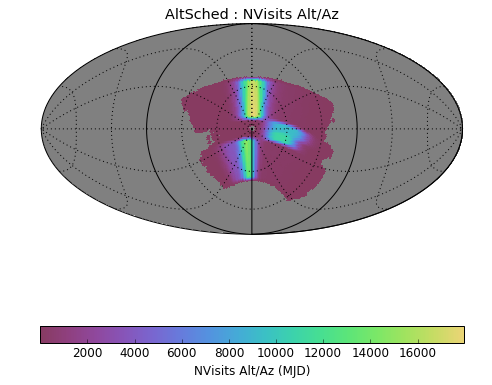}
    }
    \caption{Number of visits as a function of alt/az (North is up,
             East is right, zenith is center, and the horizon is shown as a bold
             line). \minion exhibits an East bias,
             where observations are preferentially taken at high
             airmass in the East. \altsched stays close to the meridian
             except near azimuth, where LSST's alt/az mount prevents
             observations directly on the meridian.}
    \label{fig:altaz}
\end{figure}

Maximizing $T_{eff}$ is one motivation for \altsched's meridian-scanning strategy, since
observing fields at their minimum airmass yields the highest $5\sigma$ depth. Results for
these three metrics are shown in Table \ref{table:total_teff}. Despite achieving a lower
$OSF$ and higher average slew time, \altsched reaches approximately the same $T_{eff}$
as \minion, since \minion observes off the meridian, as shown in Figure \ref{fig:altaz}.
In particular, in LSST's wide-fast-deep region, \minion achieves a mean (median)
airmass of 1.22 (1.21) compared to \altsched's 1.12 (1.09). \minion's mean (median)
normalized airmass -- i.e. the airmass of an observation divided by the minimum
airmass that field could have been observed at -- is 1.16 (1.14) compared to
1.05 (1.01) for \altsched.
\altsched suffers from a higher slew time for three reasons: first, because
we change filters much more often than \minion in order to obtain same-night colors
for nearly every visit; second, because our scanning strategy is simple and could be
optimized for faster slews; and third, because our sky tiling is spaced farther apart
than the tiling used in \opsim.

\begin{table}
    \centering
    \begin{tabular}{l|lll}
                  & $T_{eff}$ & $OSF$ & Avg. Slew \\ 
        \minion   & 333 days & 0.72 & 7.4\footnote{The average slew
            time reported elsewhere for \minion is 6.8 seconds. However, the
            current version of the LSST software stack, which we use for \altsched,
            produces a value of 7.4, which is the value that is comparable to the 11.1
            seconds we report for \altsched.} \\
        \altsched & 329 days & 0.69 & 11.1 
    \end{tabular}
    \caption{$T_{eff}$, open-shutter fraction, and average slew time of \minion
             and \altsched. \altsched matches \minion's $T_{eff}$ despite higher
             slew times because we observe on the meridian, boosting SNR of each
             observation.}
    \label{table:total_teff}
\end{table}

\begin{figure}
    \centering
    \includegraphics[width=250px]{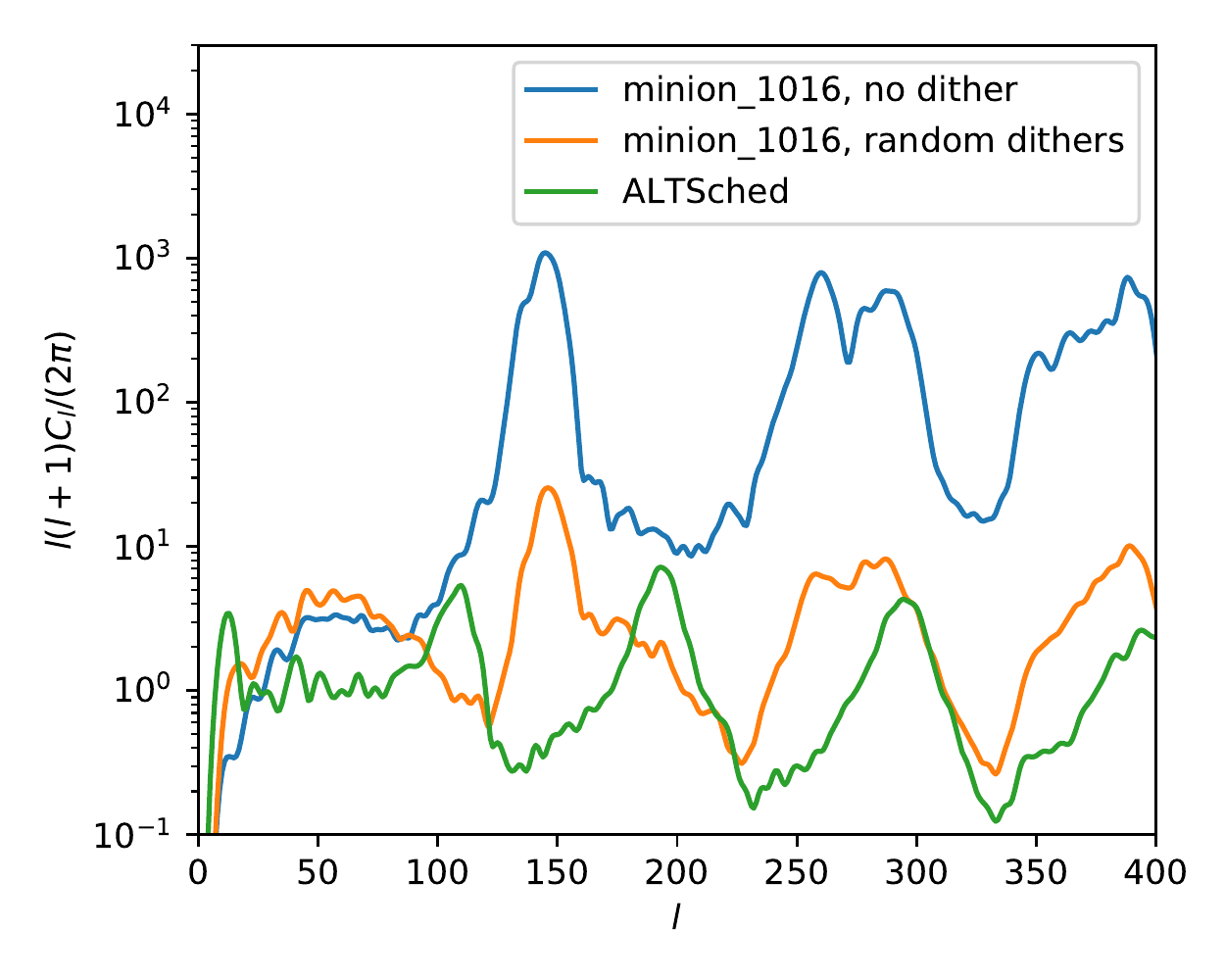}
    \caption{Angular power spectra of the number of visits to a sky pixel for a region
             of LSST's wide-fast-deep region that excludes any deep drilling fields
             and the galactic plane. Even after applying random dithers to \opsim's
             fixed fields, \altsched increases uniformity over \minion at most angular
             scales.}
    \label{fig:uniformity}
\end{figure}

Another important consideration for static science is that of survey uniformity.
In particular, weak lensing and large-scale structure measurements depend sensitively
on the uniformity of co-added depth across the sky \citep{awan+2016}.
\opsim uses a set of fixed field centers, and attempts to mitigate the resulting
imprint on the co-added depth maps by dithering around those field centers.
However, every dithering strategy tried thus far still leaves a discernible imprint
on the final co-added depth maps. Instead of dithering, \altsched eliminates fixed
fields entirely. We use a fixed tiling for the entire survey, but every night, we randomly
rotate this tiling by much more than a field size. Pointings not visited in a night are
scheduled in a subsequent night, so gaps in the tiling pattern do not persist.
To measure survey uniformity, we include angular power spectra of the number of visits
to a sky pixel, shown in Figure \ref{fig:uniformity}. \altsched's tiling strategy
reduces power at most angular scales, including around $\ell\approx 150$,
where cosmological probes are particularly sensitive.

\begin{figure}
    \centering
    \includegraphics[width=350px]{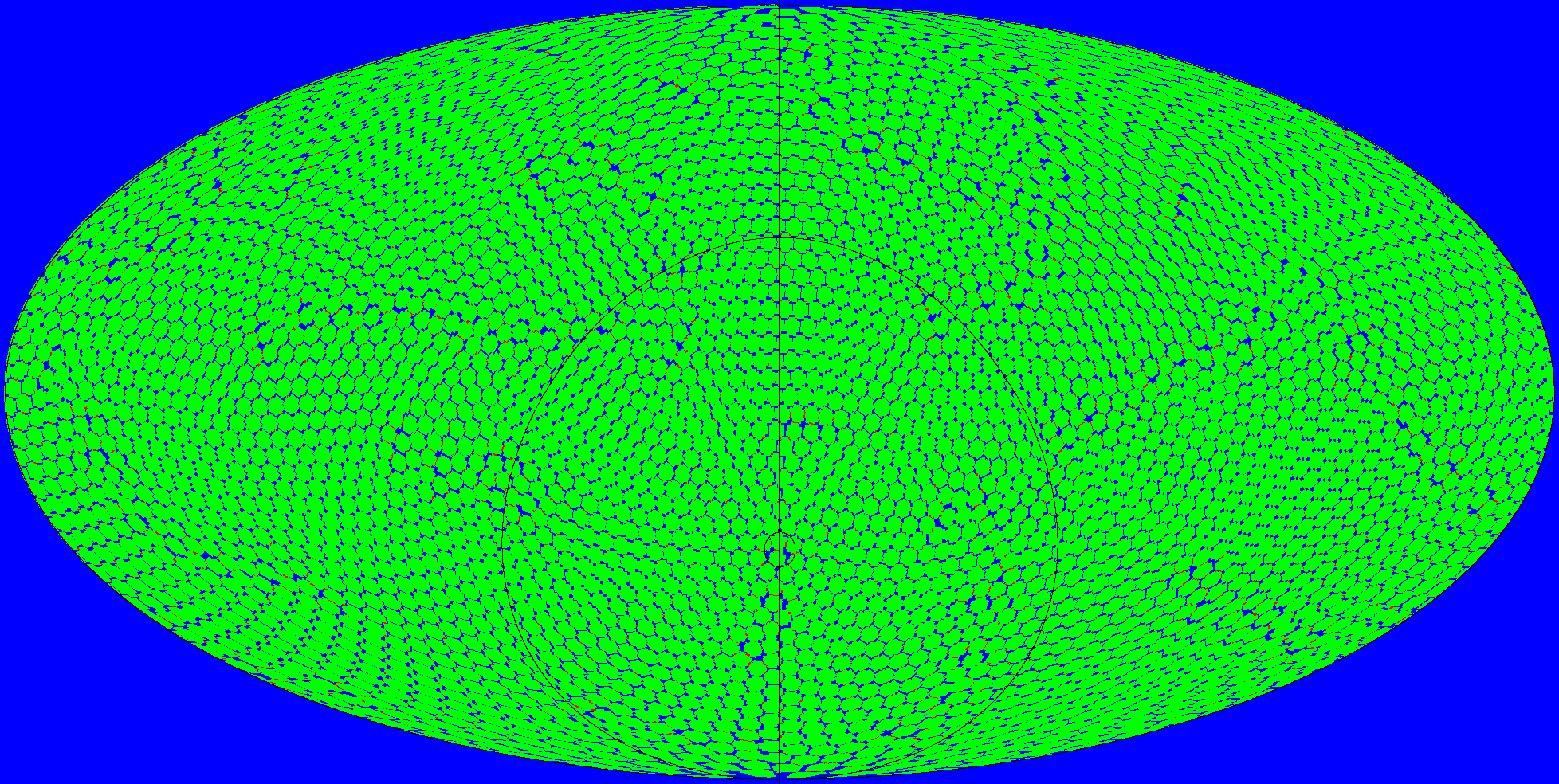}
    \caption{Tiling used by \altsched projected using the Mollweide (equal-area) projection.
             Blue indicates no visits to that sky pixel, green
             indicates a single visit, and red indicates two visits. Note that, since the
             tiling is randomly rotated each night, the gaps in the tiling do not persist
             across nights.}
    \label{fig:tiling}
\end{figure}

Unlike any strategy using fixed fields with dithering, our method admits simple analysis that
yields an expression for the uniformity in number of visits to a sky pixel. For the
fixed tiling used throughout the survey, let $\Omega_{0,1,2}$ be the areas of sky covered
by 0, 1, and 2 pointings, respectively. Assume no sky pixels are observed more than twice,
so the total area covered by the tiling $\Omega=\Omega_0+\Omega_1+\Omega_2$. Then the
standard deviation of number of visits to a sky pixel using our strategy -- i.e.
after applying the randomly-rotated tiling $n$ times -- is
\[\sigma = \sqrt{n\left(\frac{\Omega_1+4\Omega_2}{\Omega}-\left(\frac{\Omega_1+2\Omega_2}{\Omega}\right)^2\right)}.\]
See Appendix \ref{app:uniformity} for a brief derivation. Note that this
expression does not depend on the tiling itself -- only on the $\Omega_i$. In
order to minimize slew time while maintaining an even cadence, we choose a
tiling with the fields evenly spaced and with $\Omega_2 \approx 0$.
Even with $\Omega_2 \approx 0$, we achieve some fast revisits due to pointings
held over from previous nights, which will overlap randomly with the current
night's pointings. To make the fields ``evenly spaced'', we draw pointings from a
solution to the Thomson charge-distribution problem for $N=3500$ \cite{thomson1904}.
See Figure \ref{fig:tiling} for a visualization of our tiling.
Adjusting $N$ changes the density of the tiling, which controls the tradeoff
between the frequency of rapid revisits and the area observed per night.

\subsection{Time-Domain Science}
Optimizing time-domain science is much more about managing tradeoffs between different
science cases than maximizing certain quantities.
We argue in this section that
1) the parameters of \altsched directly control these tradeoffs, whereas in \opsim,
the tradeoffs are difficult or impossible to control, and
2) the \altsched simulation analyzed in this paper enables a wider range of science
cases than \minion. 

Most existing simulations assume that LSST will carry out pairs of visits separated
by $\sim 30-60$ minutes (to link asteroid observations), and will take 30-second exposures
during each visit. Some \opsim simulations have been run to explore deviating from
these assumptions, but here we take them as given. In this section, we explore
how the distribution of visits over time affects time-domain science tradeoffs,
and describe how these tradeoffs can be controlled in \altsched and \opsim.

One fundamental tradeoff in time-domain science is between the mean season length
and the mean inter-night gap (or mean cadence), which is the mean duration between
consecutive visits to a sky pixel. Controlling this tradeoff with \altsched is simple:
instead of scanning along the meridian, the scheduler can start the night either East
(shorter season/higher mean cadence) or West (longer season/lower mean cadence) of the
meridian, and slowly move West/East over the course of the night. In contrast, there is no
direct way to control this tradeoff with a greedy scheduler like \opsim. The simplest way
would be to relax the airmass/hour angle penalty and hope that the scheduler makes
full use of the additional area it can use. However, this is exactly what \minion does, and
instead of increasing the season length, it simply observes at a higher, but mostly
fixed, airmass (see Figure \ref{fig:altaz}). Alternatively, one could adjust the penalty
on the hour angle throughout the night in order to persuade the scheduler to start
observing in, say, the East, and then move over to the West by the end of the night.
This might achieve the desired result, but would require tuning of weights just in
order to reproduce the simple behavior already achieved by \altsched.

\begin{figure}
    \centerline{
        \includegraphics[width=220px]{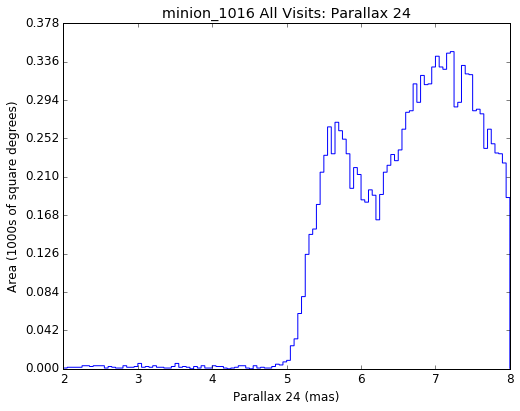}%
        \includegraphics[width=220px]{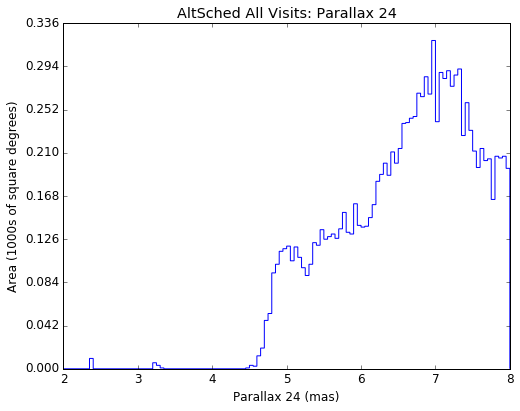}
    }
    \caption{Parallax precision for an $r=24$ magnitude star (without refraction),
             using either \minion (left) or \altsched (right). Lower is better.}
    \label{fig:parallax}
\end{figure}

In practice, we observe along the meridian throughout the night, since we don't see
an advantage in changing the season length/mean cadence at the expense of lowering SNR
by observing at higher airmass. Because \minion observes at a wider hour angle range,
and therefore has a slightly longer season length, one might worry that \altsched suffers
in parallax performance. However, this is not the case, as shown in Figure \ref{fig:parallax}.

\begin{figure}
    \centerline{\includegraphics[width=220px]{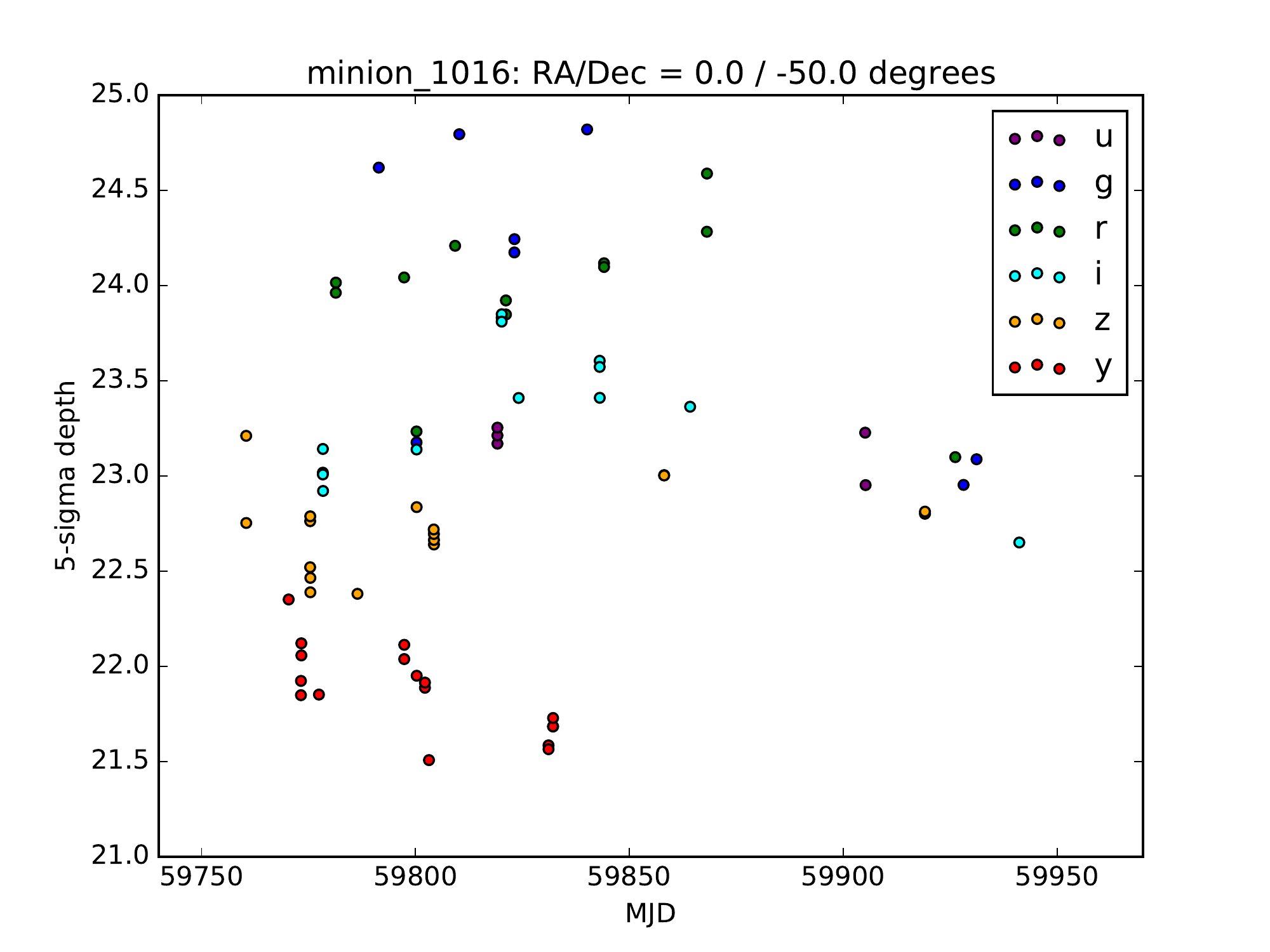}%
                \includegraphics[width=220px]{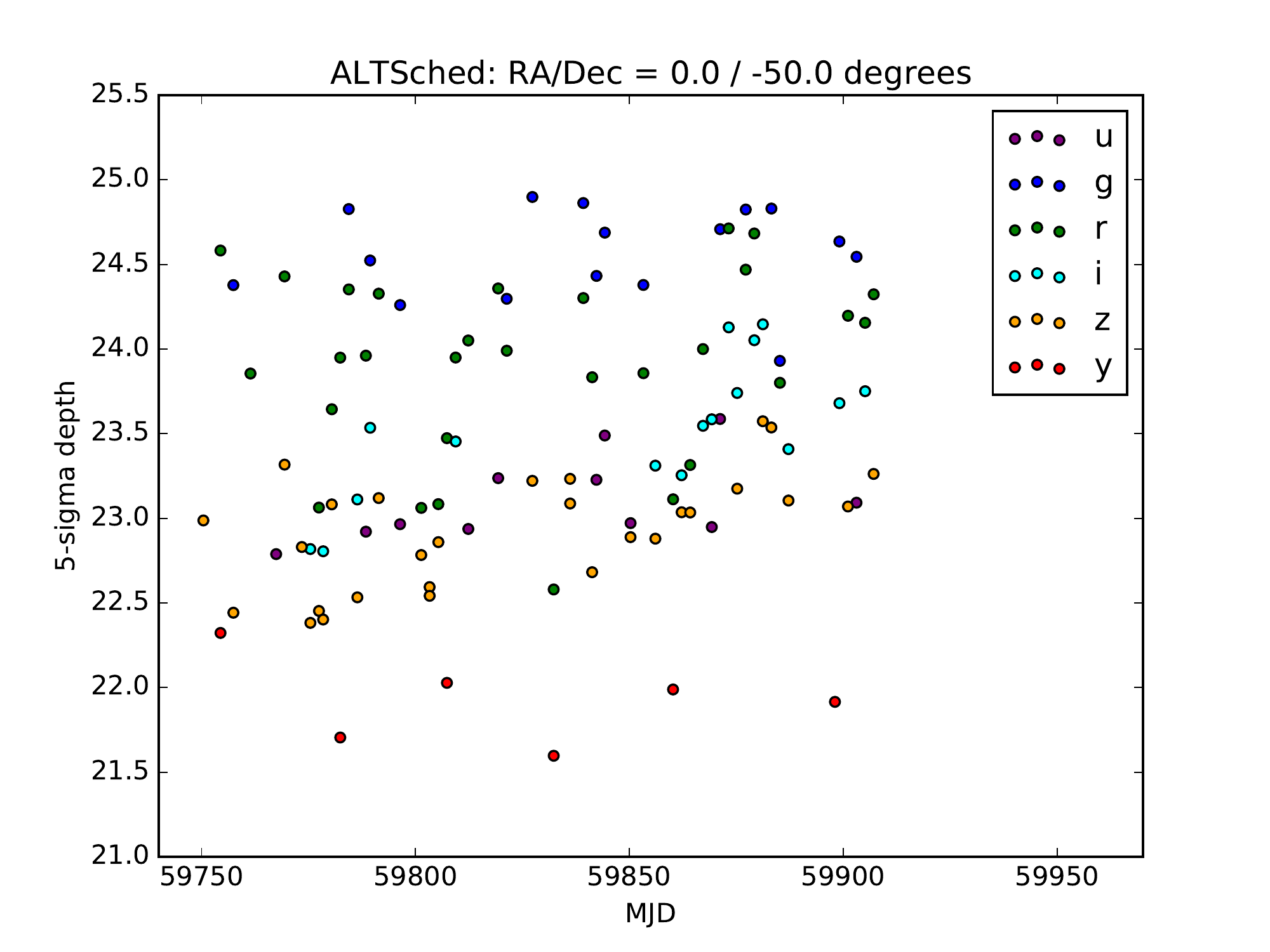}}
    \caption{$5\sigma$ depth vs time for an arbitrarily chosen RA/Dec for
             \minion (left) and \altsched (right). These plots are typical
             for each scheduler for LSST's wide-fast-deep region. The two plots
             have approximately the same number of visits (94 for \minion vs.
             97 for \altsched), but \altsched spreads the visits more uniformly
             over time.}
    \label{fig:time_plots}
\end{figure}

\begin{figure}
    \centerline{
        \includegraphics[width=220px]{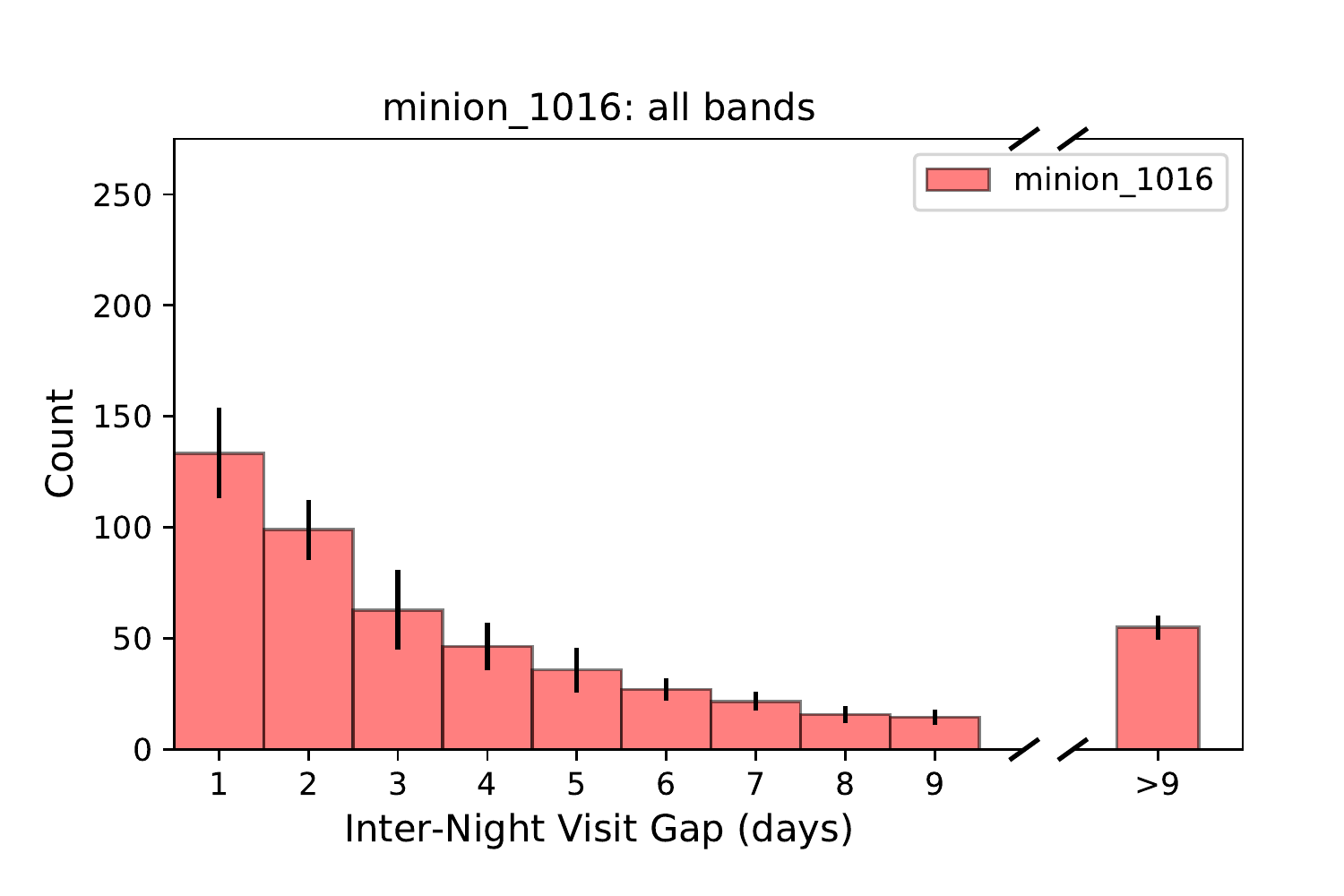}%
        \includegraphics[width=220px]{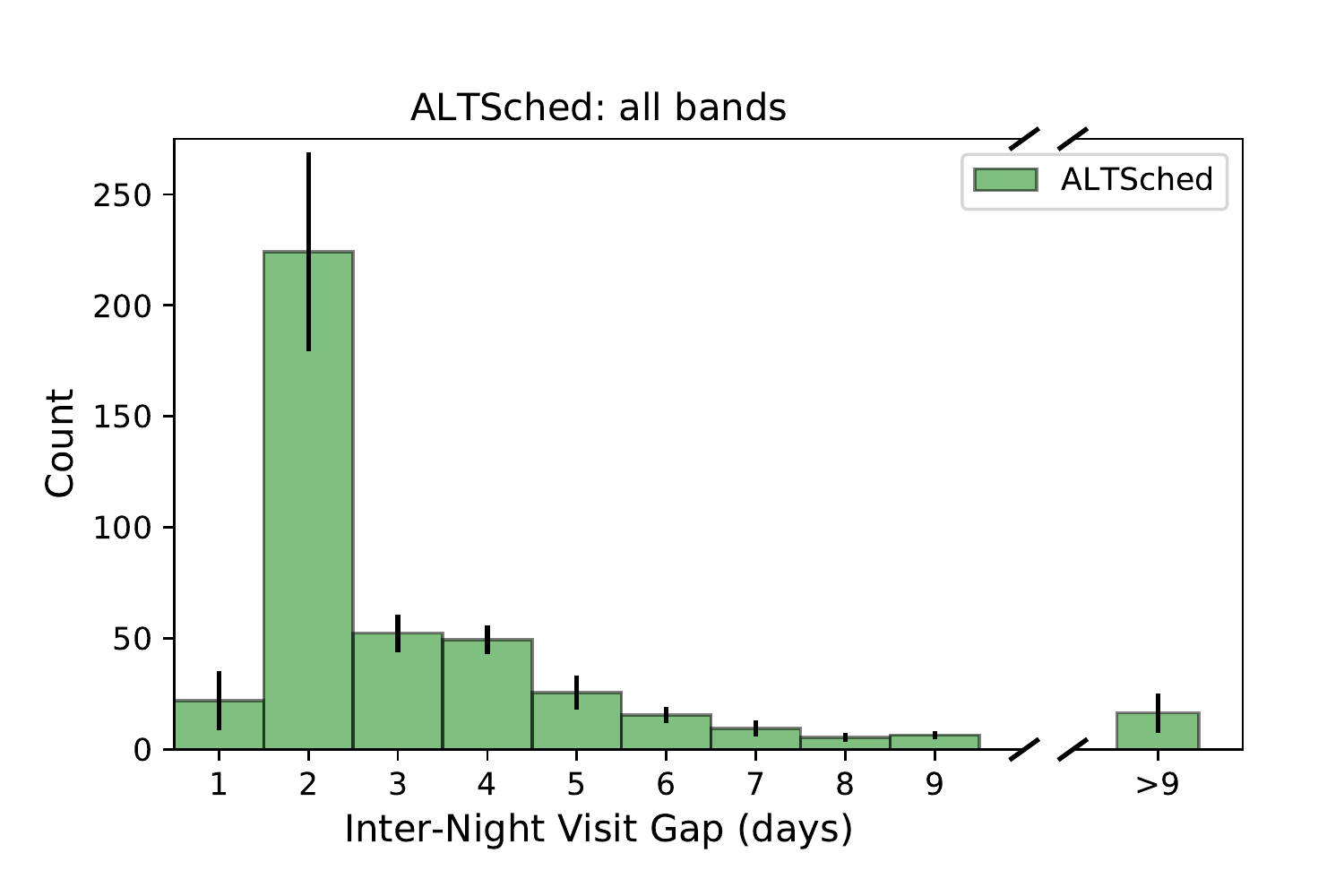}
    }
    \caption{Histogram of inter-night visit gaps to a sky pixel in
             \minion and in \altsched, for visits in any filter.
             The mean inter-night gap is roughly conserved, since both schedulers
             have a similar total number of visits. But by alternating between
             observing the northern and southern skies, \altsched suppresses
             1-night revisits, which significantly reduces longer gaps. The
             difference between the two schedulers is even more striking than these
             plots indicate, since most of the remaining long gaps in \altsched
             are attributable to downtime, weather, and season gaps.}
    \label{fig:inter_night_gaps}
\end{figure}

\begin{figure}
    \vspace{-40px}
    \centerline{
        \includegraphics[width=220px]{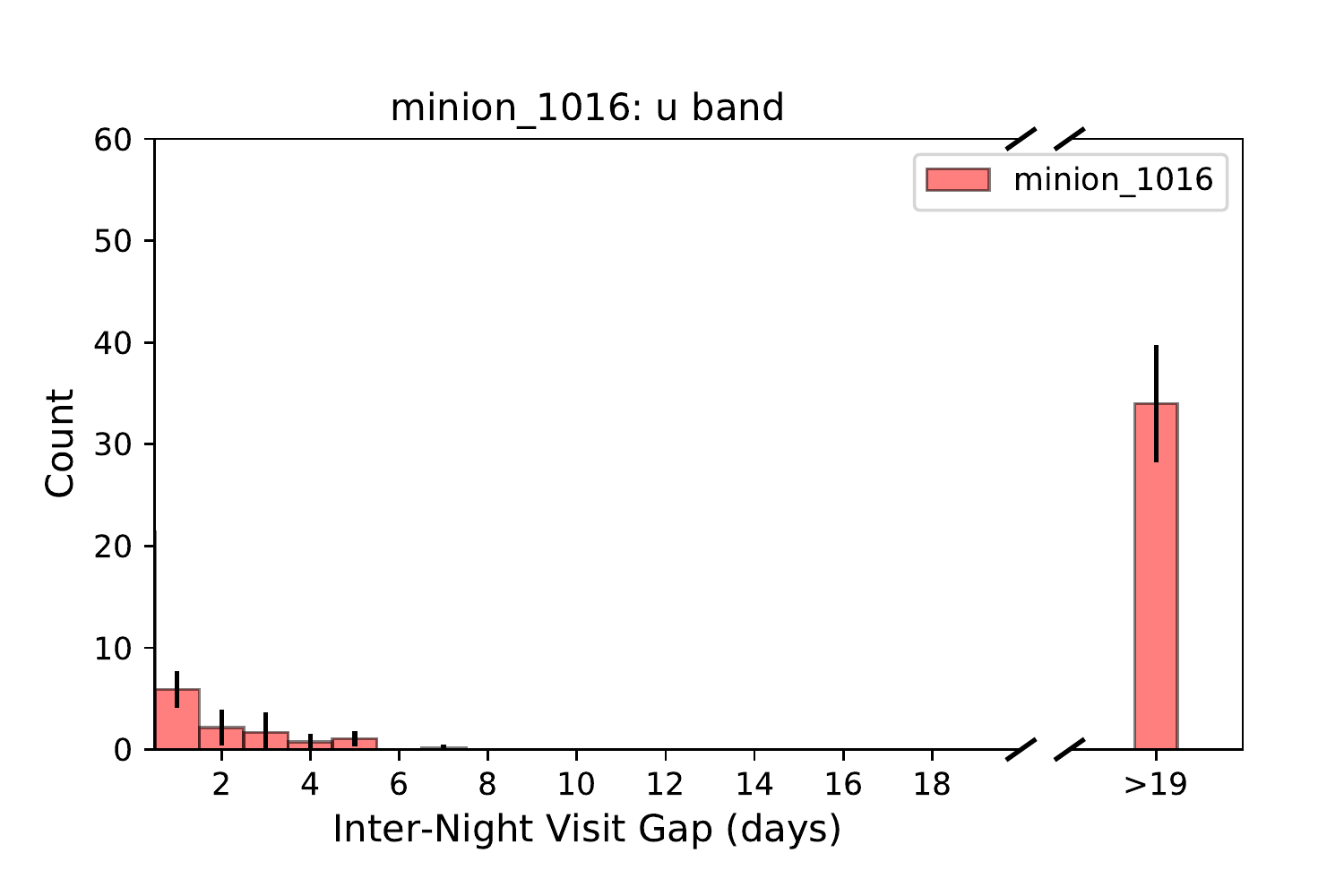}%
        \includegraphics[width=220px]{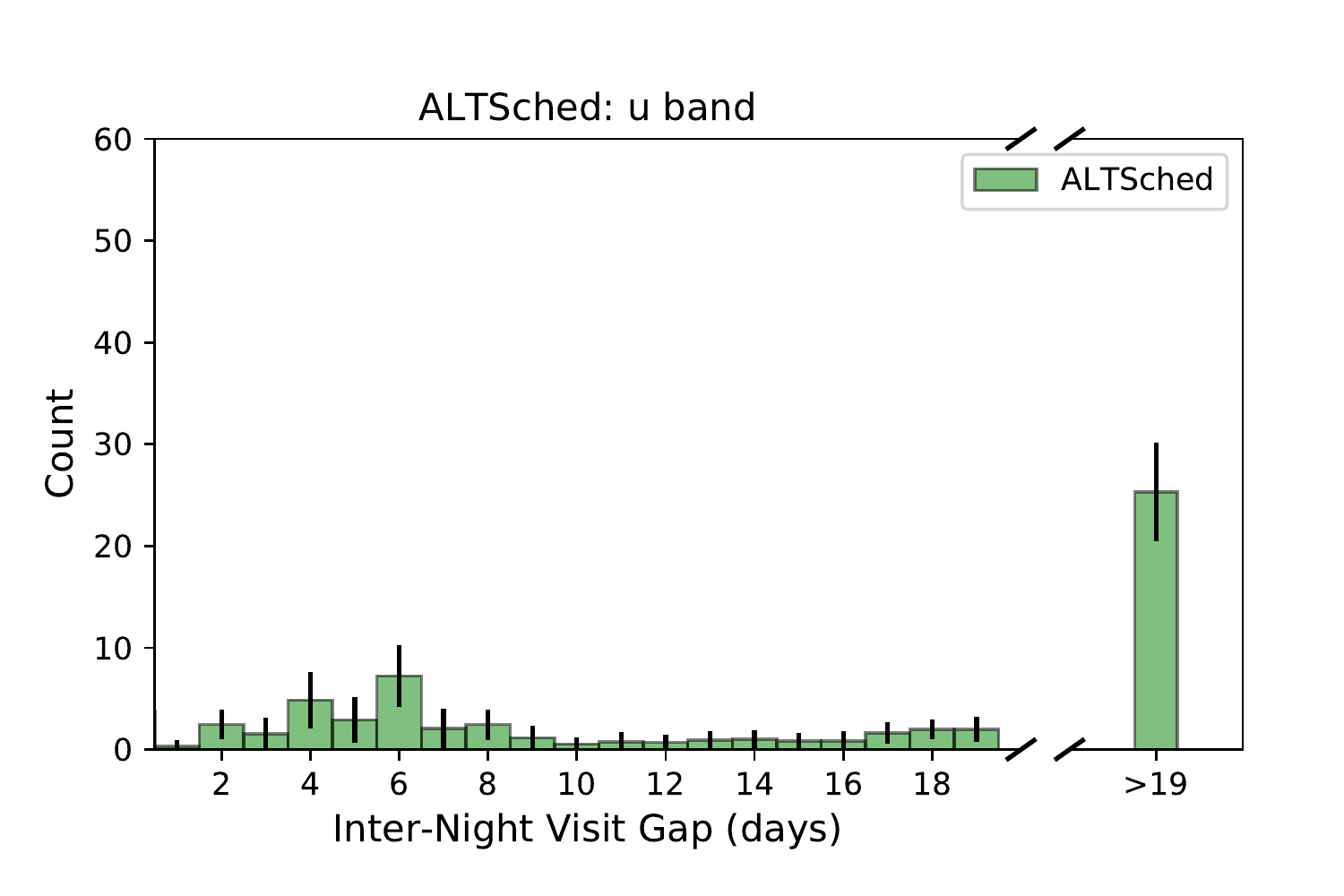}
    }
    \centerline{
       \includegraphics[width=220px]{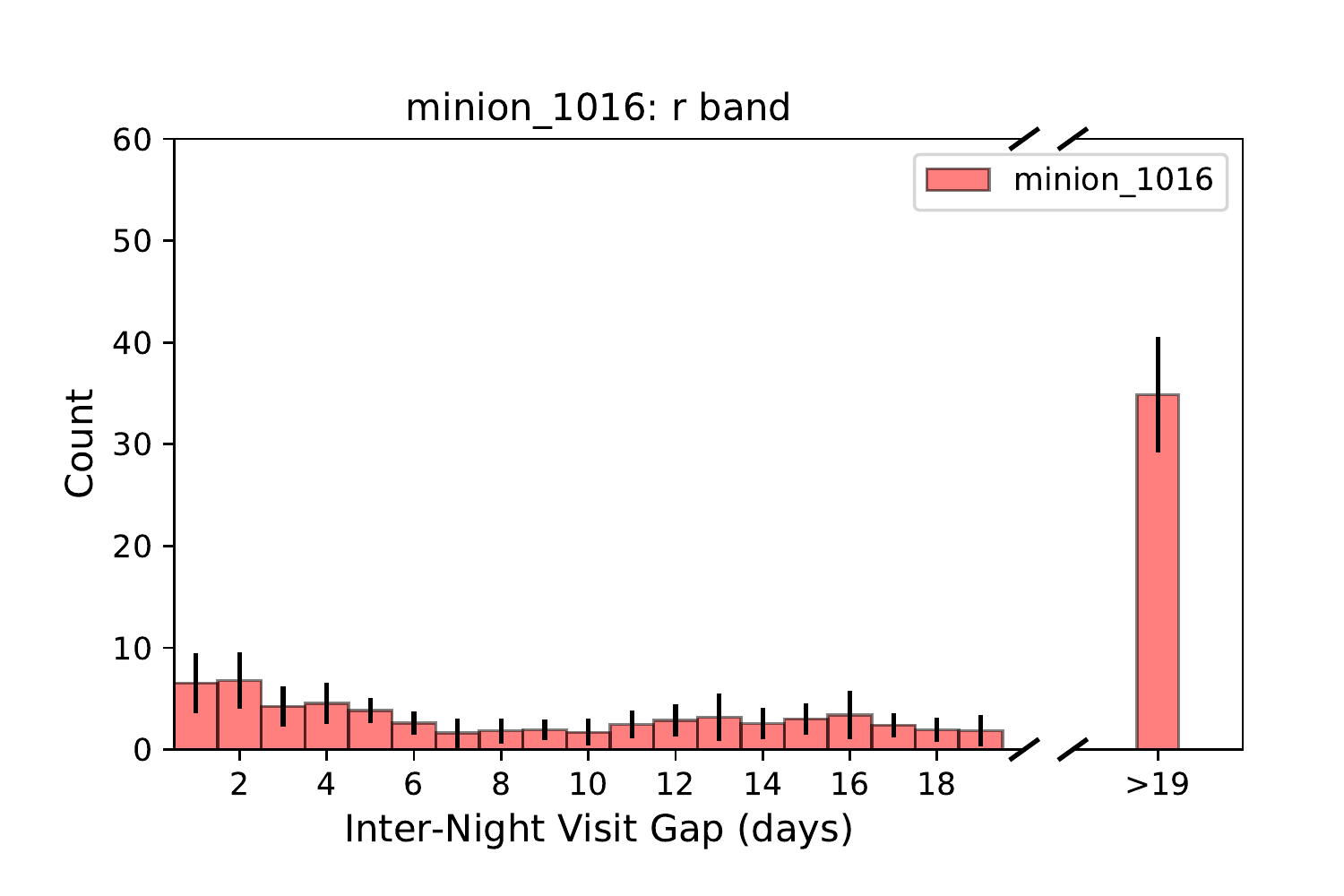}%
       \includegraphics[width=220px]{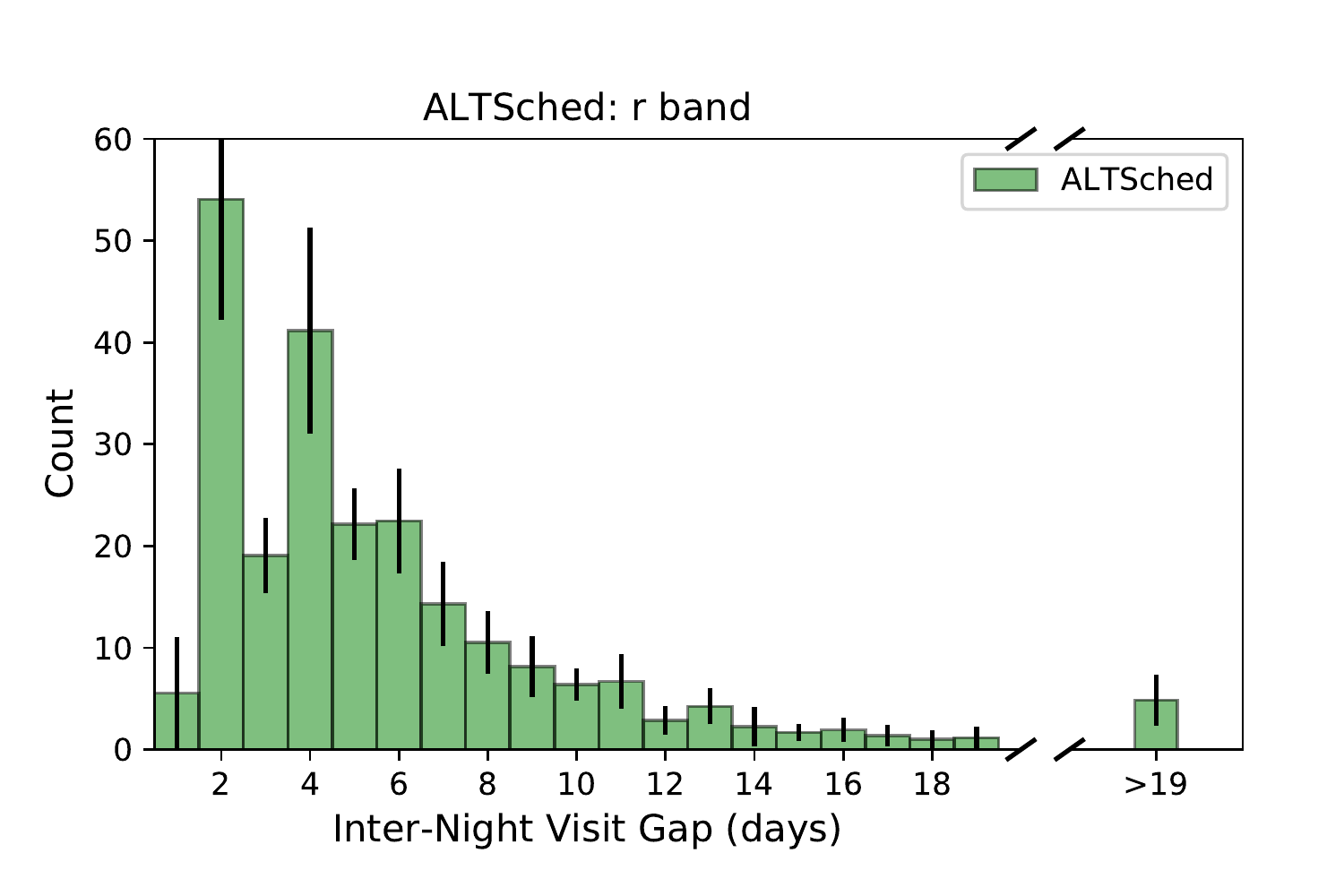}
    }
    \centerline{
        \includegraphics[width=220px]{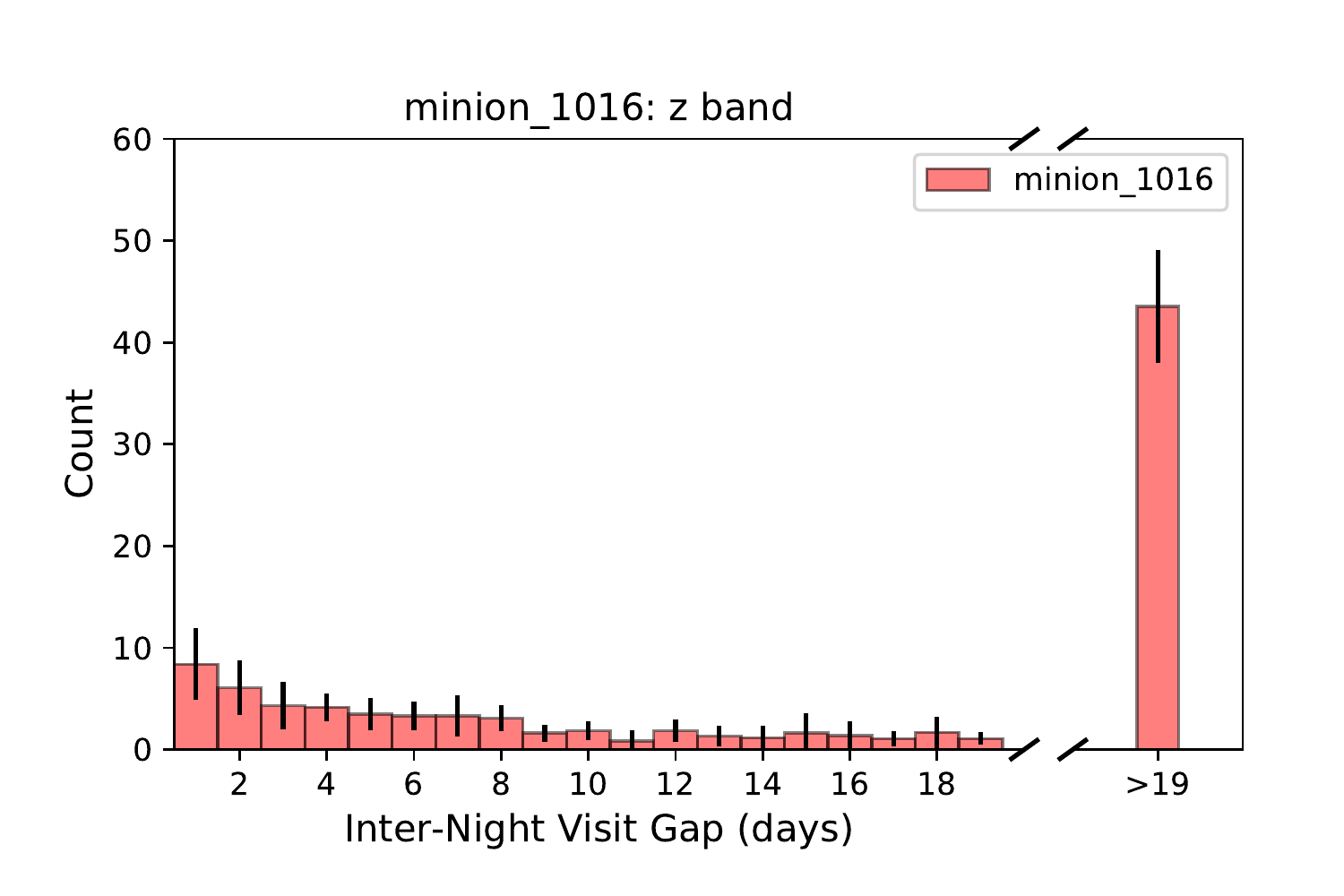}%
        \includegraphics[width=220px]{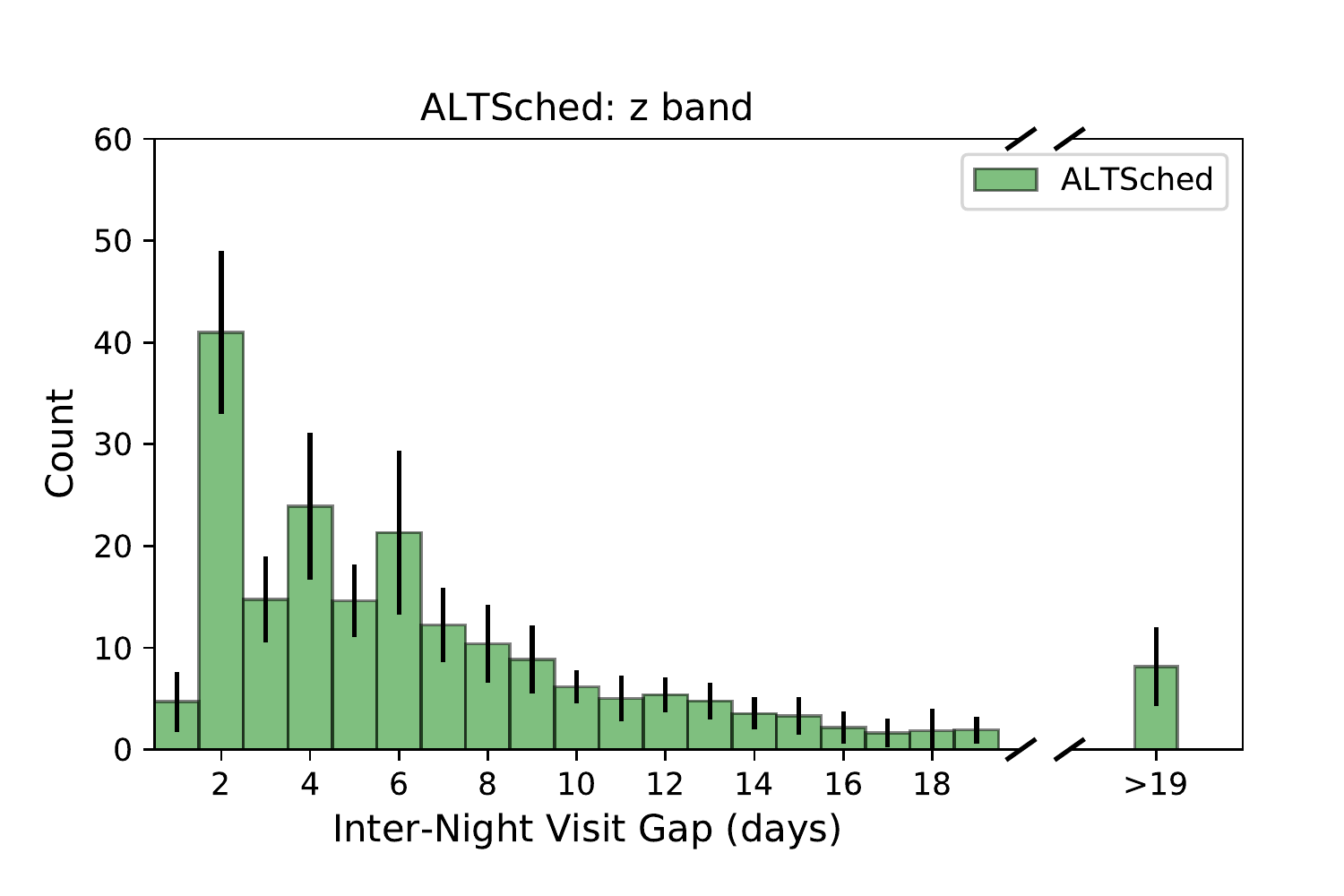}
    }
    \caption{Per-band inter-night visit gap histograms for \minion
             and \altsched, for the $u$, $r$, and $z$ bands. Since
             both \altsched and \minion defer $u$ observations to dark time, we only
             see a small gain relative to \minion in the $u$ band. Histograms for
             $g$, $i$, and $y$ are similar, and are deferred to Appendix \ref{app:gaps}.}
    \label{fig:inter_night_gaps_urz}
\end{figure}

Although the mean inter-night gap is largely determined by observing efficiency and the
season length, as described above, the actual distribution of inter-night gaps,
measured with an \textit{inter-night gap histogram}, is much more sensitive to the
scheduling algorithm. This metric, though often disregarded, has a large impact
on the quality of the light-curves LSST will measure. To get an intuitive sense
for how the scheduling algorithm can affect the inter-night gap histogram,
consider Figure \ref{fig:time_plots}, which shows, for a randomly chosen sky
pixel, the cadences achieved by \minion and \altsched. Notice that \minion often
observes this pixel many times in quick succession, and then goes many days without
any re-observations. In contrast, \altsched observes this pixel with a much
more regular cadence. This intuition is quantified with histograms of the gaps between
consecutive visits to a sky pixel, shown in Figures \ref{fig:inter_night_gaps} and 
\ref{fig:inter_night_gaps_urz}.

Controlling the inter-night gap histogram with \altsched is also simple. Changing the
mean season length/mean cadence as described above yields a simple scaling of the histogram
in the x-axis. And the location of the peak can be tuned either by changing the number of 
visits to a field per night, or by employing a rolling cadence. Both options 
are directly controllable in \altsched. The sharpness of the peak can also be controlled,
simply by judiciously choosing which nights the telescope observes North vs. South.
In the default version of \altsched, the scheduler observes North and South on alternating
nights, yielding a sharp peak in the inter-night gap histogram at 2 days. However, choosing
a repeating sequence such as N N S N S S N S would yield a flatter peak at 2 days, with
half the histogram mass at 2 days, and the other half distributed between 1 and 3 days.
In contrast, every inter-night gap histogram we have measured from an \opsim simulation
(even for simulations released after \minion) has looked roughly exponential. Note that
an exponential distribution of inter-night gaps is consistent with
an unconstrained stochastic process about the mean.
We infer from this that none of the tunable parameters in \opsim explored so far have
a discernible impact on the shape of the inter-night gap histogram.

Many science cases for LSST rely on approximately month-long transients (supernovae,
kilonovae, tidal disruption events, etc.). For these science cases, we expect
that multi-week long gaps will severely reduce the quality of light curves, without a
commensurate increase in quality from the higher rate of sampling over small sections of
the light curve. This intuition is borne out in simulations, run by Nicolas
Regnault, Philippe Gris, and the Paris Supernovae Cosmology Team, of the number of 
well-sampled type Ia supernovae different simulated schedules would obtain
(personal communication). As shown in Table  \ref{table:snIa}, \altsched
achieves an eightfold increase in the number of well-sampled SNeIa, and also
an increase of 0.07 in the maximum redshift at which the type-Ia  supernova
sample is complete. Although these simulations are for SNeIa in particular, we
expect similar results to hold for other transients. For example, \citet{kilonovae} find
that \altsched achieves a nearly 2x increase in the number of serendipitous kilonovae 
discoveries compared to \minion. Similarly, \citet{goldstein2018glsn} find that \altsched
discovers lensed supernovae earlier than \minion, enabling faster spectroscopic
follow-up.

\begin{table}
    \centering
    \begin{tabular}{l|cc}
                 & $N_{SNe}$ & Avg. $z_{max}$ \\
         \minion  & 47,000   & 0.30 \\
         \altsched & 366,000 & 0.37
    \end{tabular}
    \caption{Results from simulations run by Nicolas Regnault, Philippe Gris, and
             the Paris Supernova Cosmology Team (personal communication). $N_{SNe}$
             is the number of well-sampled type-Ia supernovae (SNeIa) in the
             redshift-limited sample. Here, a SN is well-sampled if the light
             curve has (considering only griz observations):
             1) $\geq 1$ visit every 4 days; 2) $>1$ visit in [-20, -10] days (restframe);
             3) $>1$ visit in [+35, +45] days (restframe); and 4) $\sigma_{color} < 0.04$.
             $z_{max}$ is the maximum redshift at which the sample
             of SNeIa is complete, and the average in ``Avg. $z_{max}$'' is taken over
             sky pixels.}
    \label{table:snIa}
\end{table}

\begin{figure}
    \centerline{
        \includegraphics[width=220px]{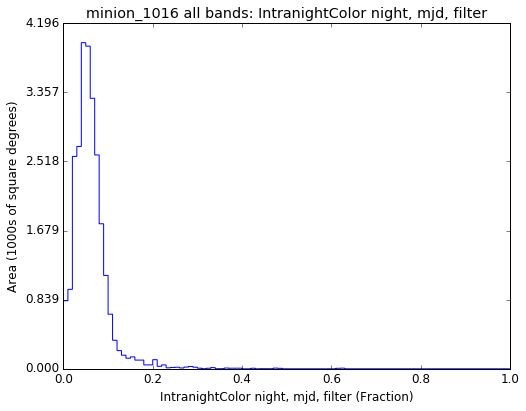}%
        \includegraphics[width=220px]{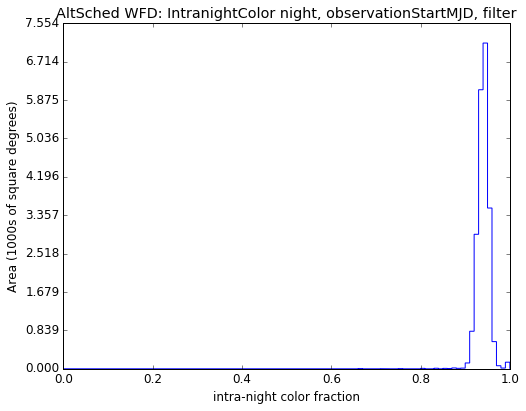}
    }
    \caption{Histogram of the fraction of nights in which a sky pixel
             was visited in at least two different bands. Only nights
             when two or more exposures were achieved at a sky pixel are
             included in the calculation. By observing nearly all fields
             in two bands instead of only one in a given night, \altsched
             doubles the observing cadence in each band.}
    \label{fig:intranight_color}
\end{figure}

\altsched achieves more favorable inter-night gap histograms than \minion
for two main reasons. The band-agnostic gaps (Figure \ref{fig:inter_night_gaps})
are improved because \altsched  alternates between observing the northern and
southern skies each night, suppressing 1-day revisit gaps. Because the mean
cadence is approximately conserved across schedulers with similar numbers of
total visits, this by necessity eliminates most of the long gaps, which are so
detrimental for transient characterization. The per-band gaps (Figure 
\ref{fig:inter_night_gaps_urz}) are improved since we carry out the two visits
each sky pixel receives per night in different filters, doubling the cadence in
each band (see Figure \ref{fig:intranight_color}).

Although in this paper we advocate looking at the entire inter-night gap histogram,
it is more common within LSST scheduling to present histograms of the median inter-night gap.
We present these histograms in Appendix \ref{app:median_gaps}.

So far, we have considered only the inter-night gaps between visits. However, the histogram
of gaps between observations to a sky pixel within a given night is also an important metric.
In order to sample transients at a large range of time-scales, both \altsched and
\minion revisit each field after $~30-60$ minutes. In addition, both schedulers
carry out some number of ``rapid revisits,'' which are spaced less than a minute apart.
\minion carries out more rapid revisits than \altsched, and the frequency can be adjusted
in either scheduler by changing the field tiling density: a tiling with more overlaps
between adjacent pointings yields more \tildemid 30-second revisits. We therefore omit
further analysis of rapid revisits.

\section{Schedulers used for other sky survey projects} \label{sec:related}

The scheduling principles we describe in this paper apply to any
ground-based telescope able to image a large sky area per unit time.
Such telescopes include LSST \citep{lsst}, Palomar Transient Factory
(PTF) \citep{rau2009ptf}, Zwicky Transient Facility (ZTF)
\citep{smith2014ztf}, Pan-STARRS1 \citep{chambers2016panstarrs},
SkyMapper \citep{skymapper}, the Dark Energy Survey (DES) \citep{des},
and others. Scheduling software has been developed for each of these
telescopes, but often only limited information about these schedulers
can be found in the literature. Broadly speaking, most schedulers
are greedy algorithms like \opsim: they choose each observation as the night
progresses based on where the telescope is currently pointing and on current
or predicted observing conditions. In general,
greedy algorithms are guaranteed to maximize merit in the long term
in only the simplest of problems, and telescope scheduling is not
one of those problems. To see why, consider a scenario where every
field north of the telescope has a merit score of, say, 5, and every
field south of the telescope has a merit score of 4. Fields directly
overhead have very low merit. If the telescope happens to be pointing in the
South and long slews are penalized, then a greedy algorithm will call
for observing in the South the entire night, even though the globally
optimal policy would be to slew to the North at the very beginning
and observe there for the rest of the night.

The schedulers used for PTF \citep{law2009ptf}, PanSTARRS1 \citep{chambers2008drm},
and, as we understand it, DES \citep{neilsen2013obstac}, are all greedy algorithms. 
Similarly to LSST, these surveys include a wide-area component which receives multiple
epochs in several bands, and we suspect that the challenges in time-domain
science faced by greedy algorithms for LSST apply to these surveys as well.
ZTF uses a different scheduling algorithm, which schedules
an entire night at a time by solving an integer program designed to maximize
observing efficiency, subject to constraints on when and how often
each field needs to be observed \citep{ztf2019}. Las Cumbres Observatory
uses a similar algorithm \citep{lco2014}.
For LSST, this approach is challenging for a few reasons: the overhead
for each observation varies considerably (a few seconds to 2 minutes);
in order to achieve a more uniform sky coverage, LSST may not use fixed
fields at all; and the large number of fields and observations per night
may render the integer program intractable to solve repeatedly during a
night when weather changes.

\section{Conclusions, and Future Directions}

As we have described in this article, \altsched gives survey designers much more
control over \textit{global} survey characteristics than a greedy algorithm. However,
certain limitations caused by our algorithm's simplicity leave room for improvement
in the more \textit{local} scheduling decisions: \altsched does not
avoid taking exposures near the moon; it does not avoid poor seeing conditions
or clouds; and it makes slews that are longer than necessary.
These problems are not insurmountable, and because they strictly worsen
\altsched's performance on metrics, we view these limitations optimistically
as opportunities to extract even more science performance from LSST.

Between now and the beginning of full operations, the LSST Project plans to find an
optimal survey strategy by running a large number of \opsim simulations, and to
choose whichever yields the best compromise between science cases.
We want to stress that the best survey strategy that can be found with this procedure
is likely to under-perform on a number of time-domain metrics, since the parameters
of \opsim don't give much control over the survey's cadence.

Instead, we advocate for combining the advantages of \altsched (global schedule
characteristics) with those of \opsim (sensible local scheduling decisions).
We see two ways to combine forces. The
simplest is to simulate surveys that switch back and forth between the two
algorithms. When \altsched tries to observe the moon, switch to a greedy algorithm
that knows not to; at the end of every month, spend a day or two using a greedy
algorithm to even up co-added depth across the sky; when clouds are present, use
the greedy algorithm exclusively. Strategies like these should
have minimal impact on the favorable properties of \altsched, while also largely
rectifying some of its limitations. Another way forward is to combine the two
algorithms directly, by using a greedy scheduler whose merit function has been
specifically engineered to reproduce \altsched for large-scale decisions (which
region of the sky to observe; which filter to use), but where more local
scheduling decisions can be made by lower-order terms in the merit function.

In parallel, we encourage the LSST science community to continue using \altsched
as a tool to demonstrate the minimum performance that LSST is capable of. We
developed most of \altsched over a few months, and many opportunities for improvement
remain: finding a scanning pattern that minimizes slew times; making more intelligent
filter choices; adapting throughout the night to avoid deviating from the meridian;
recovering more gracefully from downtime; using separate tilings per-filter to increase
homogeneity; etc. Making improvements like these, and combining \altsched with
\opsim as described above, is what we believe will find the best survey strategy
available in the limited time remaining before full operations commence.

\acknowledgments
We are thankful to numerous members of the LSST Project team for extensive
conversations about scheduling choices and system characteristics. Stubbs
and Rothchild acknowledge support from the US Department of Energy under
grant DE-SC0007881, and from Harvard University.
Yoachim acknowledges funding from the LSST Corporation.

%

\vspace{5mm}


\software{astropy \citep{robitaille2013astropy},
          Metrics Analysis Framework  \citep{jones2014lsst}
          }




\newpage
\bibliography{altsched}

\newpage

\begin{appendices}
\section{Median Inter-Night Gaps}
\label{app:median_gaps}

\begin{figure}[h]
    \centerline{
        \includegraphics[width=220px]{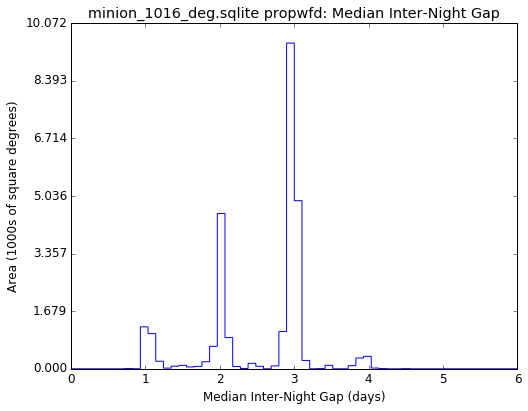}%
        \includegraphics[width=220px]{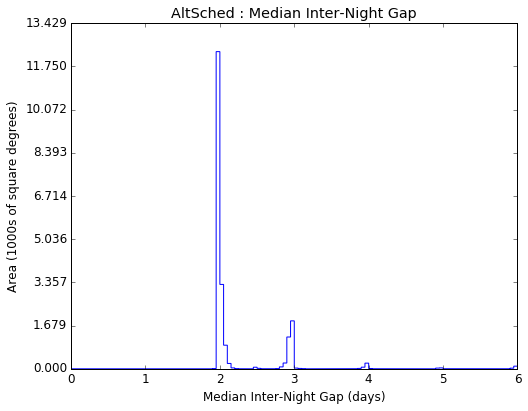}
    }
    \caption{Histogram of median inter-night visit gaps to a sky pixel in
             \minion and in \altsched, for visits in any filter.
             \minion has slightly longer
             observing gaps because it effectively has a longer season
             duration and because it observes some fields more than twice
             per night.}
    \label{fig:median_gaps}
\end{figure}

\begin{figure}
    \vspace{-40px}
    \centerline{
        \includegraphics[width=220px]{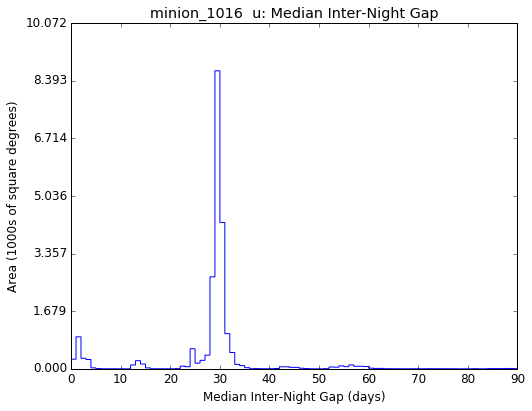}%
        \includegraphics[width=220px]{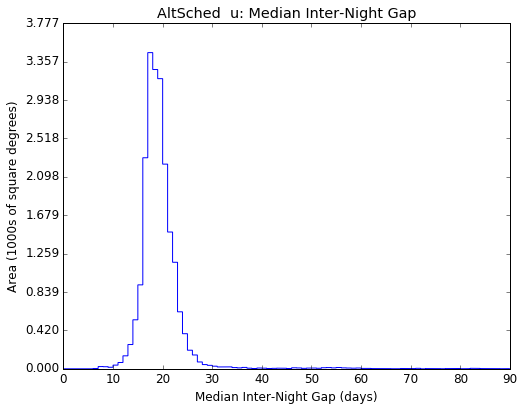}
    }
    \centerline{
       \includegraphics[width=220px]{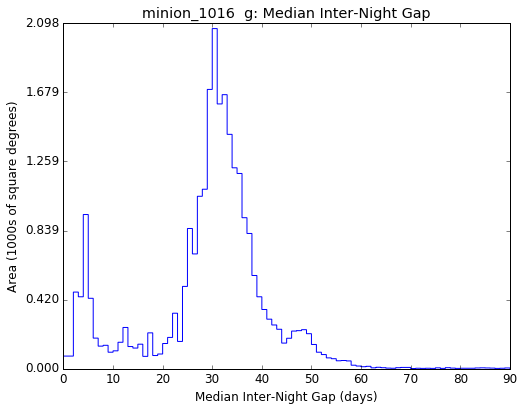}%
       \includegraphics[width=220px]{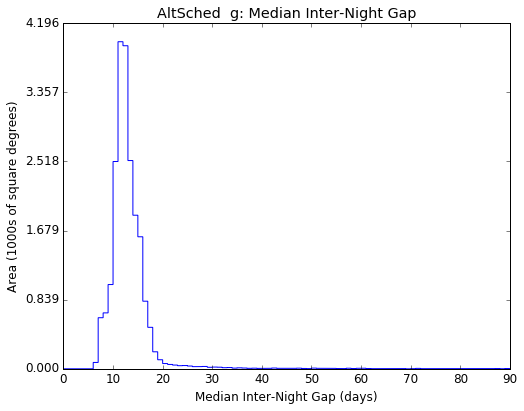}
    }
    \centerline{
        \includegraphics[width=220px]{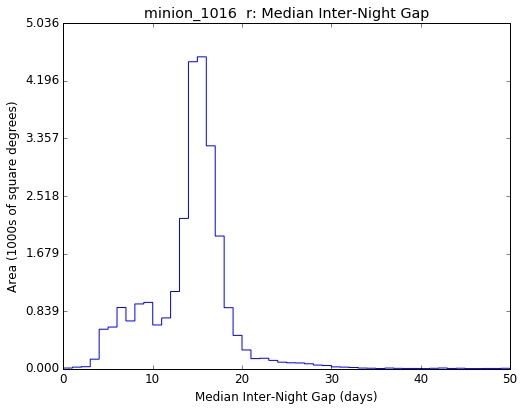}%
        \includegraphics[width=220px]{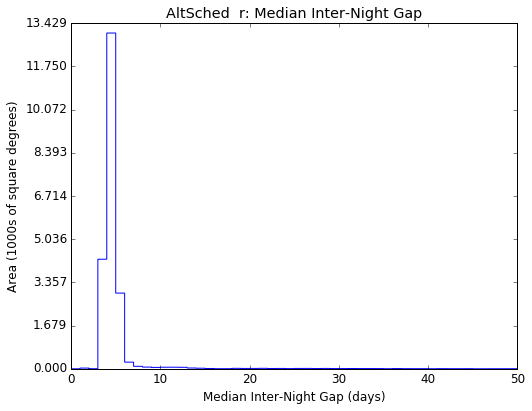}
    }
    \caption{Per-band median inter-night visit gaps for \minion
         and \altsched, for the u, g, and r bands. \altsched achieves
         considerably better median gaps, primarily because it
         executes pairs of observations taken in a single night in
         different filters, effectively doubling the per-band cadence
         in each band.}
    \label{fig:median_gaps_ugr}
\end{figure}

\begin{figure}
    \vspace{-40px}
    \centerline{
       \includegraphics[width=220px]{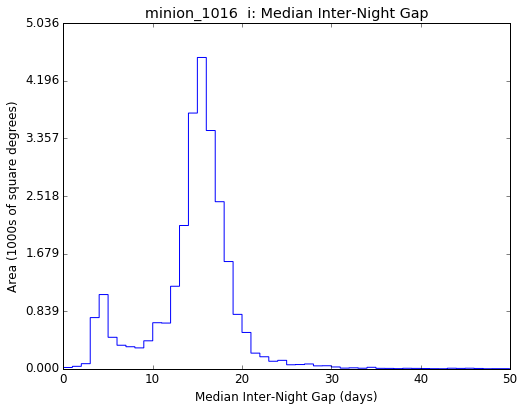}%
       \includegraphics[width=220px]{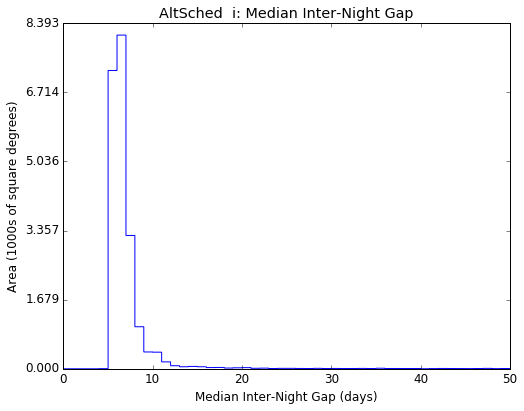}
    }
    \centerline{
        \includegraphics[width=220px]{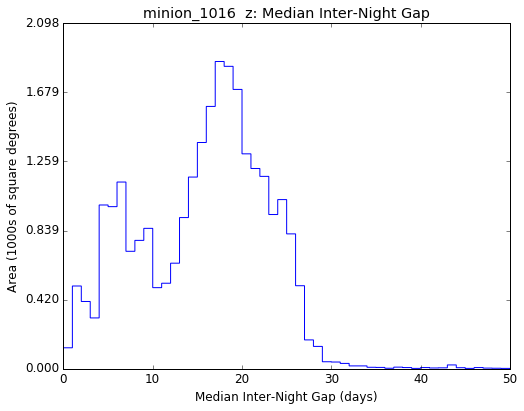}%
        \includegraphics[width=220px]{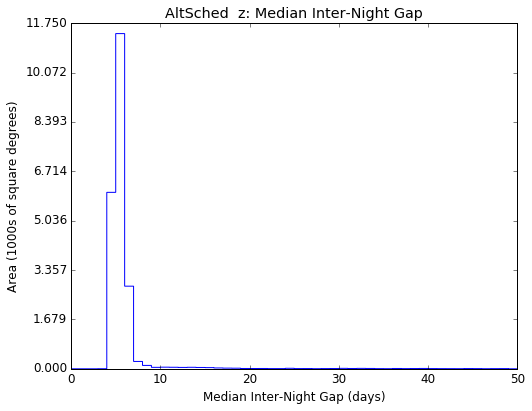}
    }
    \centerline{
       \includegraphics[width=220px]{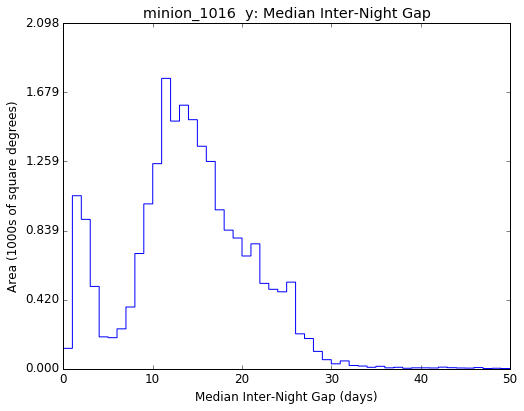}%
       \includegraphics[width=220px]{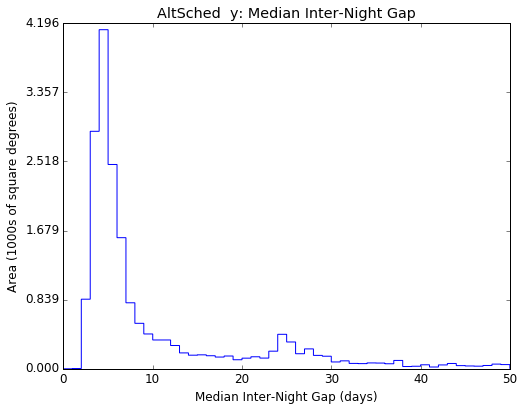}
    }
    \caption{Same as Figure \ref{fig:median_gaps_ugr}, but for the i, z, and y bands.}
    \label{fig:median_gaps_izy}
\end{figure}

Here we present histograms of the median inter-night gap between consecutive visits to 
a sky pixel. Because \minion and \altsched have a similar total number of visits
available per unit area, the mean (and therefore, roughly speaking, the median)
inter-night gap should be fixed (Figure \ref{fig:median_gaps}). We expect a two-fold reduction
in the per-band median inter-night gaps (Figures \ref{fig:median_gaps_ugr} \& 
\ref{fig:median_gaps_izy}) because \altsched observes each field in two filters per night
instead of \minion's single filter.

We see additional improvement beyond these predictions because
\minion sometimes observes the same field more than twice per night, thus
incurring additional long inter-night gaps. \minion also has a slightly longer season
length since it observes at a wider range of airmasses. For the u band, the improvement
is actually less than twofold because both \altsched and \minion cluster observations in the
u band around times with low lunar brightness.


\section{Inter-Night Gaps}
\begin{figure}
    \vspace{-40px}
    \centerline{
        \includegraphics[width=220px]{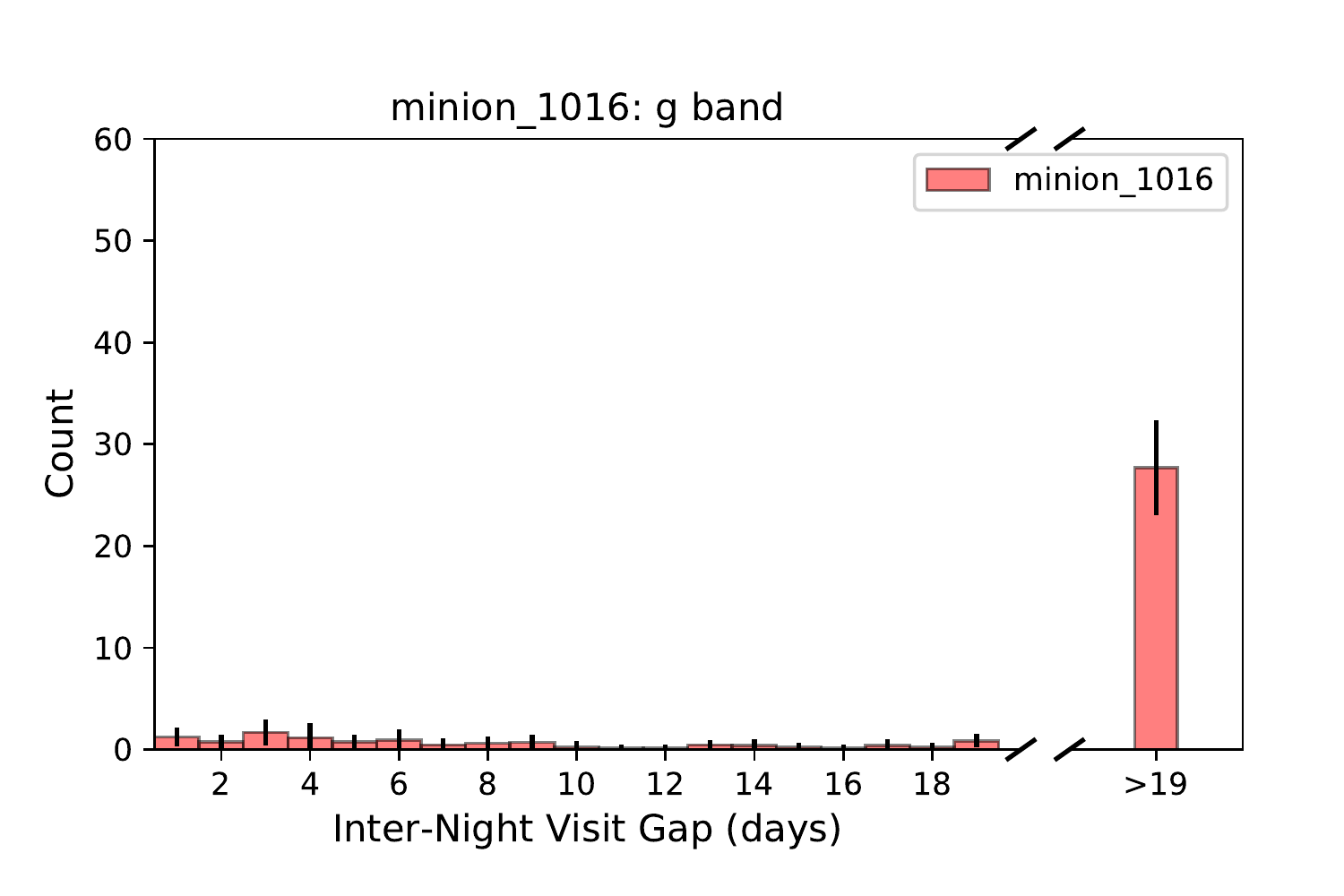}%
        \includegraphics[width=220px]{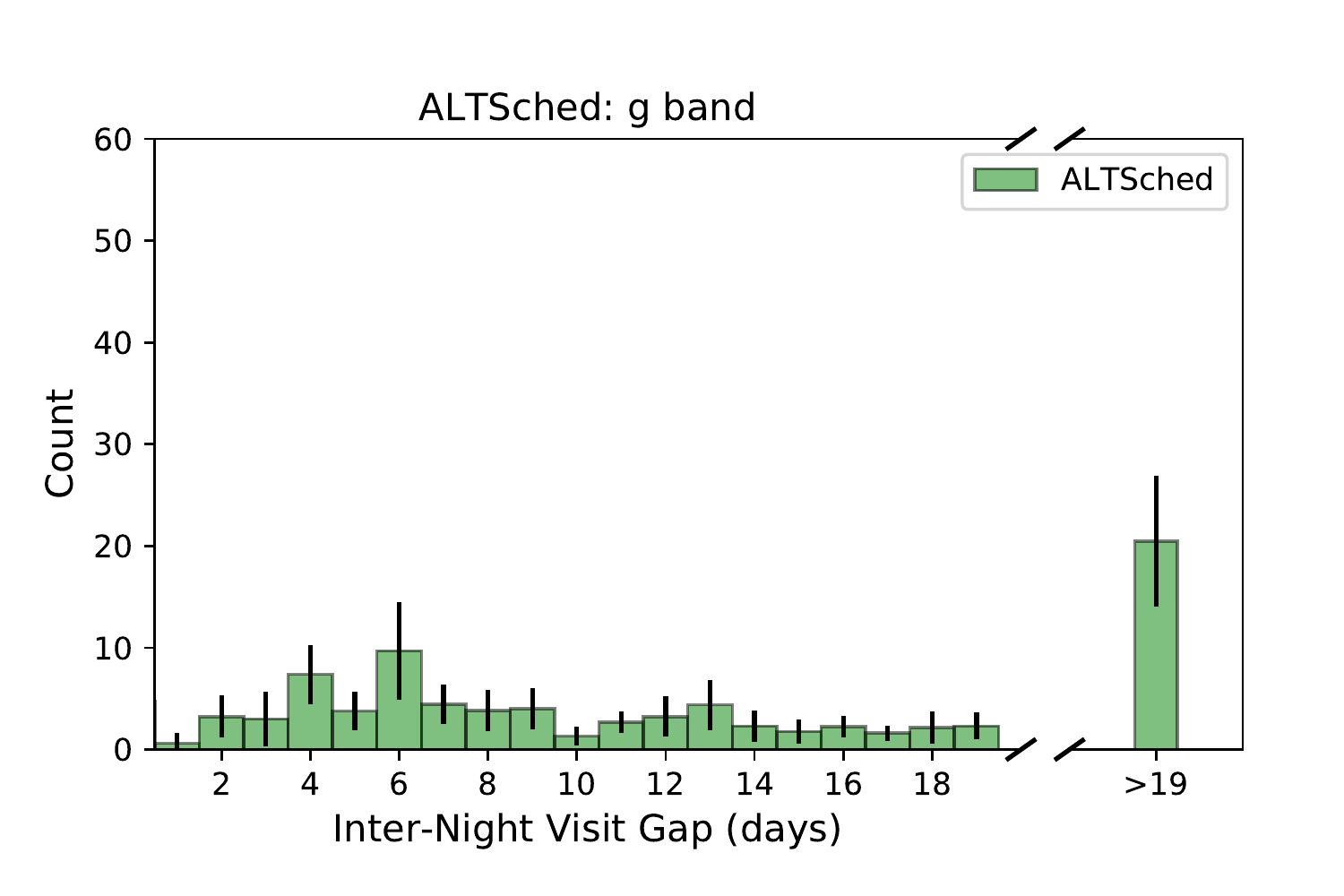}
    }
    \centerline{
       \includegraphics[width=220px]{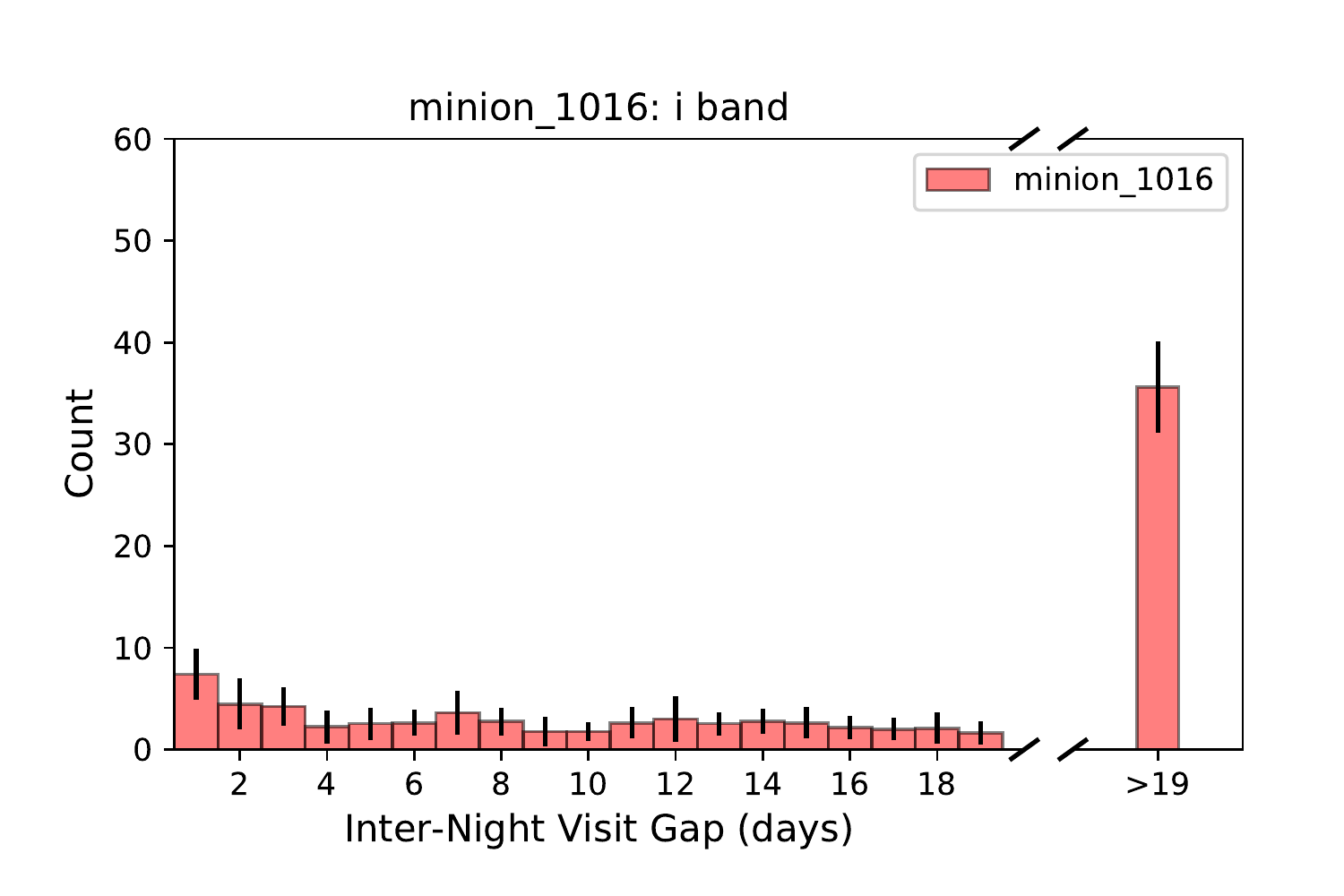}%
       \includegraphics[width=220px]{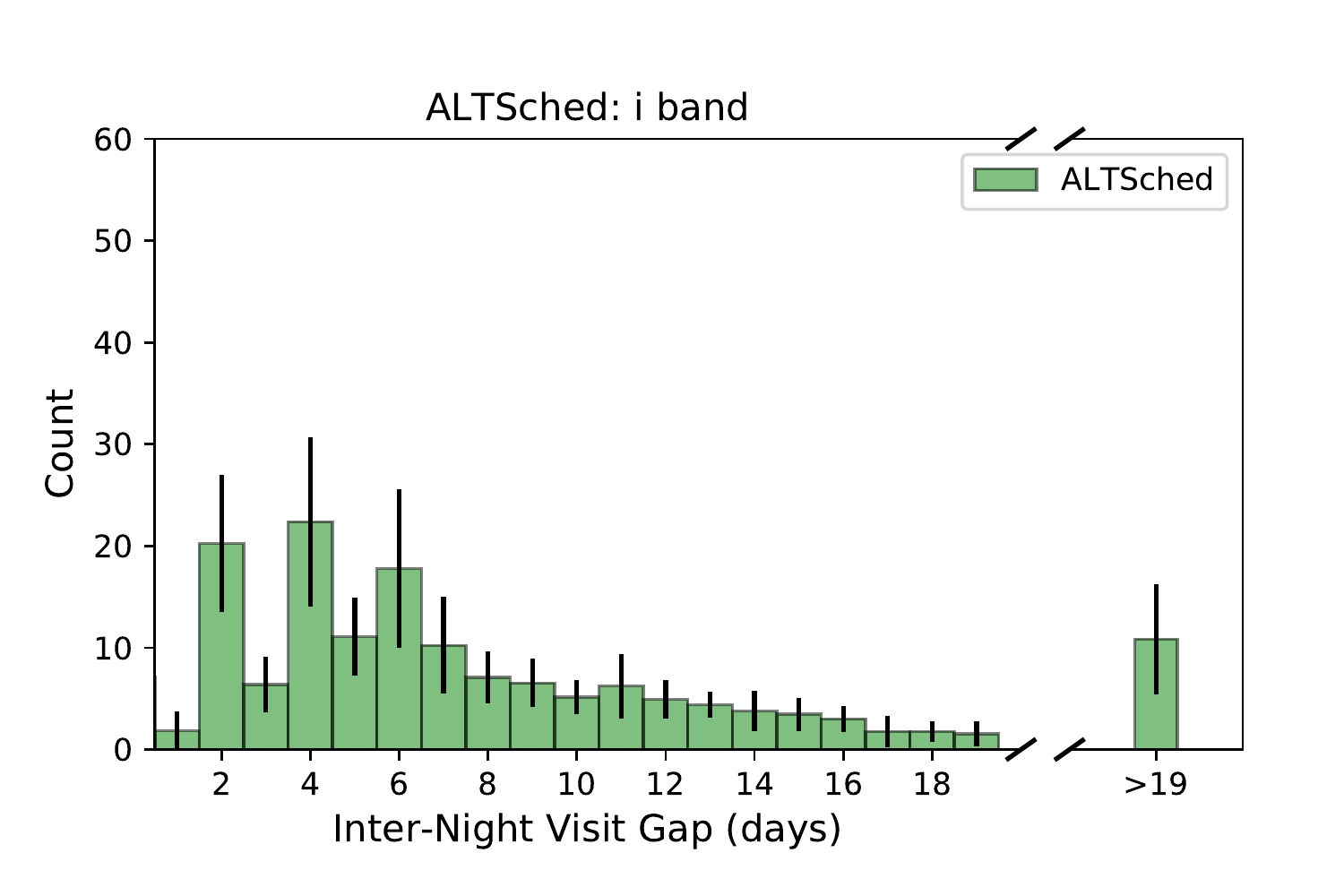}
    }
    \centerline{
        \includegraphics[width=220px]{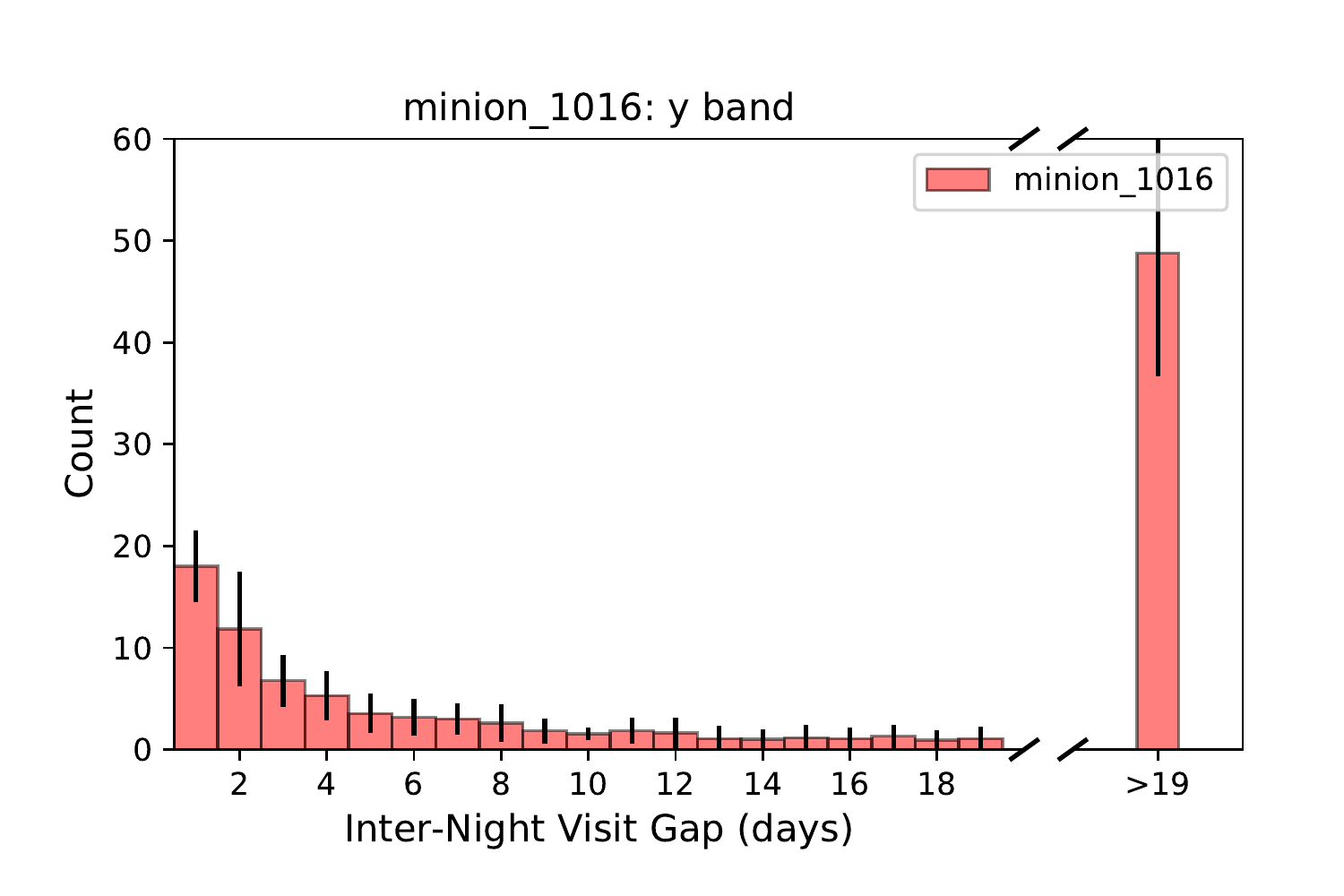}%
        \includegraphics[width=220px]{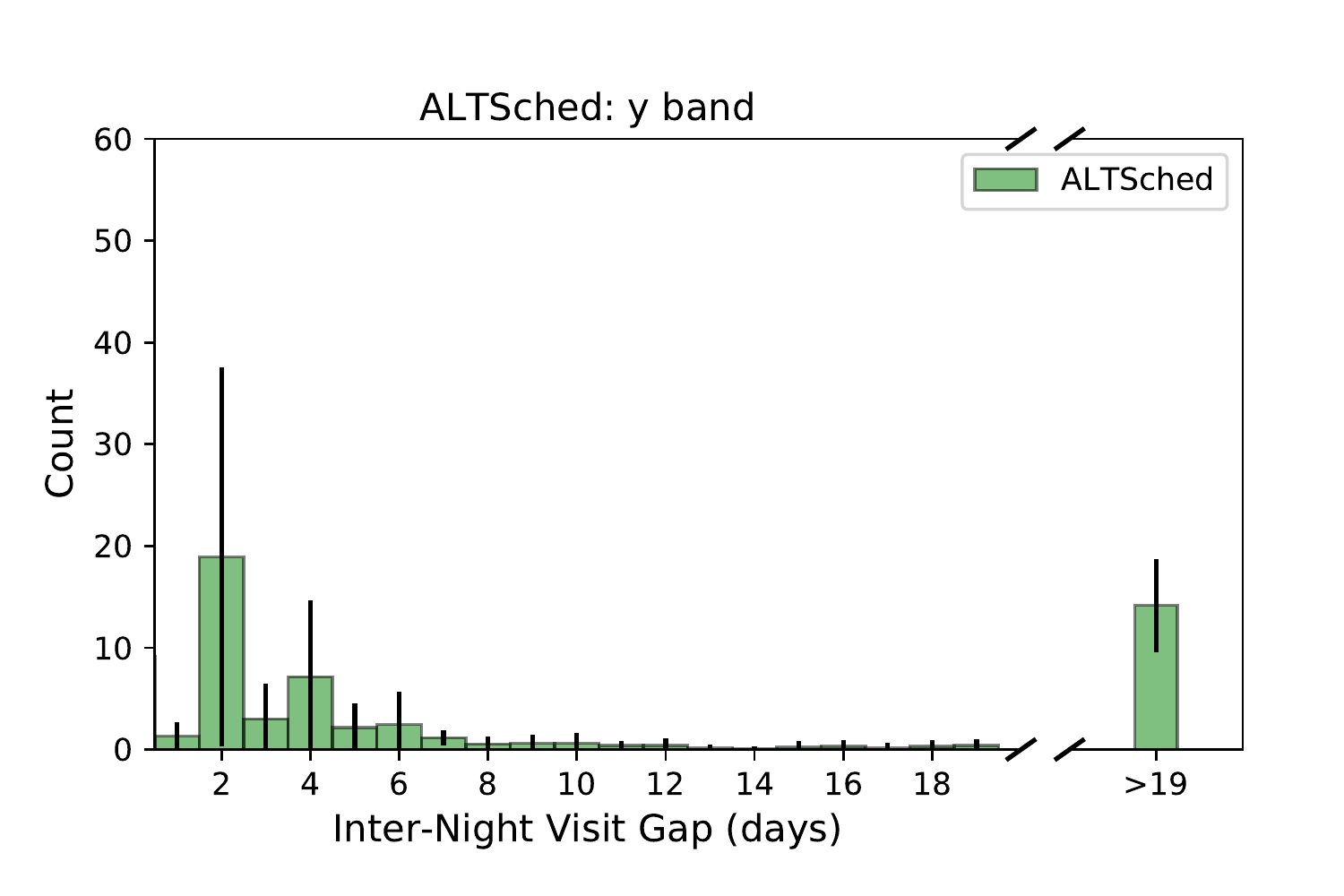}
    }
    \caption{Per-band inter-night visit gap histograms for \minion
             and \altsched, for the $g$, $i$, and $y$ bands.}
         
    \label{fig:inter_night_gaps_giy}
\end{figure}

In Figure \ref{fig:inter_night_gaps_giy}, we include histograms of the typical
inter-night gap between observations to a sky pixel in the $g$, $i$, and $y$
bands (deferred from the text).
\label{app:gaps}

\section{Uniformity Derivation}
\label{app:uniformity}
Consider a survey that observes only at pointings drawn from some fixed sky
tiling. Let the area not covered by any pointing be $\Omega_0$, the area
covered by exactly one pointing be $\Omega_1$, and the area covered by
exactly two pointings be $\Omega_2$. Assume no part of the sky is covered
by three or more fields -- i.e. that the total sky area
$\Omega=\Omega_0+\Omega_1+\Omega_2$. If this tiling is observed once, the
mean number of times a sky pixel will be observed is
\[\mu_f=\frac{\Omega_1+2\Omega_2}{\Omega}\]
and the RMS fluctuation in number of visits to a pixel is
\begin{eqnarray*}
\sigma_f & = & \sqrt{\frac{1}{\Omega}\left(\Omega_0(\mu_f)^2+\Omega_1(1-\mu_f)^2+\Omega_2(2-\mu_f)^2\right)} \\
 & = & \sqrt{\frac{\Omega_1+4\Omega_2}{\Omega}-\left(\frac{\Omega_1+2\Omega_2}{\Omega}\right)^2}.
\end{eqnarray*}

If the tiling is observed $N$ times without rotation or dithering, the
final average and standard deviation in number of visits will be $N\mu_f$
and $N\sigma_f$.

To reduce the standard deviation, we can simply rotate the tiling by some random amount
each time it is observed.
If we do this $N$ times, then by the central
limit theorem, the probability distribution of number of visits to each sky
pixel is normal with mean
\[\mu_r=N\mu_f=N\left(\frac{\Omega_1+2\Omega_2}{\Omega}\right)\]
and standard deviation
\[\sigma_r = N\frac{\sigma_f}{\sqrt{N}}=\sigma_f\sqrt{N}=\sqrt{N\left(\frac{\Omega_1+4\Omega_2}{\Omega}-\left(\frac{\Omega_1+2\Omega_2}{\Omega}\right)^2\right)}.\]

\end{appendices}

\end{document}